\documentclass[]{article}

\usepackage{color}
\usepackage{graphicx}
\usepackage{rotating}
\usepackage{multirow}
\usepackage{array} 
\usepackage{tabularx}

\usepackage[T1]{fontenc}
\usepackage[colorlinks=true]{hyperref}% add hypertext capabilities
\usepackage{dcolumn}% Align table columns on decimal point
\usepackage{bm}% bold math
\usepackage{amsmath,amssymb,amsthm} 
\usepackage{makeidx}
\usepackage{amsfonts}
\usepackage{comment}
\usepackage[usenames,dvipsnames]{pstricks}
\usepackage{subfigure}
\usepackage{braket}
\usepackage{float}
\usepackage[top=1in, bottom=1in, left=1in, right=1in]{geometry}

\usepackage[utf8]{inputenc} %unicode support
\newcommand{\vecs}[1]{\mbox{\boldmath $#1$}}
\begin{document}

\title{Graph-based data clustering via multiscale community detection}

\author{Zijing Liu\textit{$^{1}$},
Mauricio Barahona$^{1,2\ast}$}
\date{}
\maketitle

\begin{abstract} % abstract
%Recent advances in dynamics on complex networks provide a framework for multiscale community detection. 
We present a graph-theoretical approach to data clustering, which combines the creation of a graph from the data with Markov Stability, a multiscale community detection framework. We show how the multiscale capabilities of the method allow the estimation of the number of clusters, as well as alleviating the sensitivity to the parameters in graph construction.
%We apply methods from network analysis to the problem of clustering in data mining. We employ Markov Stability, a multiscale community detection framework, to perform graph-based data clustering, and show that network analysis across scales provides additional robustness that alleviates the sensitivity of clustering outcomes to the details of graph construction and allows the number of clusters to be found in an unsupervised manner.
%We show that Markov Stability is able to find the significant number of clusters automatically and unveil the multiscale structure of the data.
%The multi-resolution analysis of the graph is robust to the parameter setting in graph construction, which has dramatic influences on graph-based clustering. 
We use both synthetic and benchmark real datasets to compare and evaluate several graph construction methods and clustering algorithms, and show that multiscale graph-based clustering achieves improved performance compared to popular clustering methods without the need to set externally the number of clusters.

\end{abstract}

\footnotetext{\textit{$^{1}$~Department of Mathematics, Imperial College London, South Kensington Campus, London SW7 2AZ, UK}}
\footnotetext{\textit{$^{2}$~EPSRC Centre for Mathematics of Precision Healthcare, Imperial College London, London SW7 2AZ, UK}}
\footnotetext{$^{\ast}$~\textit{m.barahona@imperial.ac.uk}}
%%%%%%%%%%%%%%%%%%%%%%%%%%%%%%%%%%%%%%%%%%%

\section*{Introduction}
Clustering is a classic task in data mining, whereby input data are organised into groups (or clusters) such that data points within a group are more similar to each other than to those outside the group~\cite{xu2005survey}. Such a task is distinct from supervised (or semi-supervised) classification, where examples of the different classes are known \textit{a priori} and are used to train a computational model to assign other objects to the known groups.
Instead, clustering aims to find natural, intrinsic sub-classes in the data, without assuming \textit{a priori} the number or type of clusters. 
Indeed, a key open issue in this field is the principled determination of the number of clusters in an unsupervised manner, without the assumption of a generative model~\cite{sugar2003finding,azran2006spectral}. %\MB{Add references for this.  Should we say in one sentence that this problem is approached differently using: inference (stochastic block model); silhouette measures; PCA like measures of gaps (Gahramani's paper)...}
The obtained groups can then constitute the basis for a simpler, yet informative, representation of large, complex datasets. 
%which is a necessary step in data analysis when the number of objects is large.

Data clustering has a long history and there exist a myriad of clustering algorithms based on different principles and heuristics~\cite{jain1999data}. %\MB{Add a reference to a book or a very comprehensive review to clustering here, perhaps von Luxburg but maybe some books or additional reviews?} 
In their most basic form, many popular clustering techniques (e.g., k-means~\cite{macqueen1967some} and mixture models~\cite{dempster1977maximum}) are based on the assumption that the data follows an explicit (typically multivariate Gaussian) distribution. Clusters are then defined as the samples most likely generated from the same distribution, and learned by likelihood maximisation.
%\MB{You need to add a sentence here about how this distribution is used to carry out the clustering: assume model and then maximise some likelihood for the data, etc, something simple in words...} \MB{BTW, k-means assumes Gaussian?}
However, in real applications, the model that generates the data is unknown and the resulting data distribution may be complex. In this case of data-driven analysis, model-based clustering often yields poor results~\cite{shi2000normalized,ng2002spectral,de2005spectral,ye2016fuse}. %\MB{Add reference for failure of gaussian based clustering}

An alternative approach is provided by spectral clustering, which uses the eigenvectors of a (normalised) similarity matrix derived from the data to find relevant subgroups in the dataset~\cite{ng2002spectral,von2007tutorial}.
Spectral clustering is underpinned by results in matrix analysis (e.g., singular value decomposition), and has strong connections to model reduction, geometric projections and dimensionality reduction~\cite{von2007tutorial,schaub2019}. 
The choice of similarity measure is a crucial ingredient to the clustering performance but, as long as a similarity matrix can be computed, spectral methods provide an attractive choice for non-vector data or for data sampled from irregular and non-convex data manifolds~\cite{alpert1999spectral,dhillon2001co}. %\MB{you had `non-convex shapes' not clear that it would be understood.. is this OK? Add some references here} 
 
From a different perspective, the similarity matrix of a dataset can also be viewed as the adjacency matrix of a fully connected, weighted graph, where the nodes correspond to data points and the edge between two nodes is weighted by their similarity.
One can then apply graph-based algorithms for community detection or graph partitioning to the problem of data clustering.
Graph-based methods typically operate by searching for balanced graph cuts, sometimes invoking notions from spectral graph theory, i.e., using the spectral decomposition of the adjacency or Laplacian matrices of the graph~\cite{hagen1992new,chungspectral}.
%\MB{add reference to Fiedler and to Chung}.  
%
Spectral clustering can thus be understood as a special case of the broader class of graph-based clustering methods~\cite{schaub2019}.
%\cite{granell2012unsupervised,de2012complex,nguyen2017clustering}
Importantly, graph-based clustering is also able to reveal modular structure in graphs across levels of resolution through multiscale community detection~\cite{lambiotte2008arXiv,lambiotte2014random, delvenne2010stability}. This approach allows for the discovery of natural data clusterings of different coarseness~\cite{altuncu2019}, thus recasting the problem of finding the appropriate number of clusters to the detection of relevant scales in the graph. %\MB{What other possible advantages of graph-based methods? cleaner partitioning? Not sure if we can say more...}

Methods for graph construction usually involve a sparsification of the similarity (or distance) matrix under different heuristics (from simple thresholding to sophisticated regularisations) in order to extract a \textit{similarity graph} that preserves key properties of the dataset~\cite{cheng2009learning}. 
%\MB{Add here some references to Spielman, and perhaps a review of L1-based optimisations like graphical LASSO}.  
The representation of data through graphs has attractive characteristics, including the capability of capturing efficiently the local and global properties of the data through graph-theoretical concepts that embody naturally the notions of local neighbourhoods, paths, and global connectivity~\cite{tenenbaum2000global,beguerisse2013finding, lambiotte2014random}. The usage of graphs provides a natural links of spectral clustering with other clustering methods and allows for easy generalisation to a semi-supervised setting~\cite{dhillon2004kernel,kulis2009semi}.
Graphs also provide a means to capture the geometry of complex manifolds, a feature of interest in realistic datasets~\cite{bronstein2017}. %\MB{add other references if relevant here for graph-based representations} 
%\MB{We need to say a few things about why graphs are a good possibility... Why they are supposed to work better in some situations... I tried above  a few thoughts but you can add or rewrite this better.} 
Graph representations not only reduce the computational cost for spectral graph methods, but also allow us to use the techniques developed for complex networks as an alternative to address problems in data clustering. However, it has been shown that both the method of graph construction and the choice of method parameters (i.e., sparsity) have a strong impact on the performance of graph-based clustering methods~\cite{maier2008influence,daitch2009fitting,maier2013result,jebara2009graph}.

\begin{figure}[!ht]
    \centering
    \includegraphics[width=\textwidth]{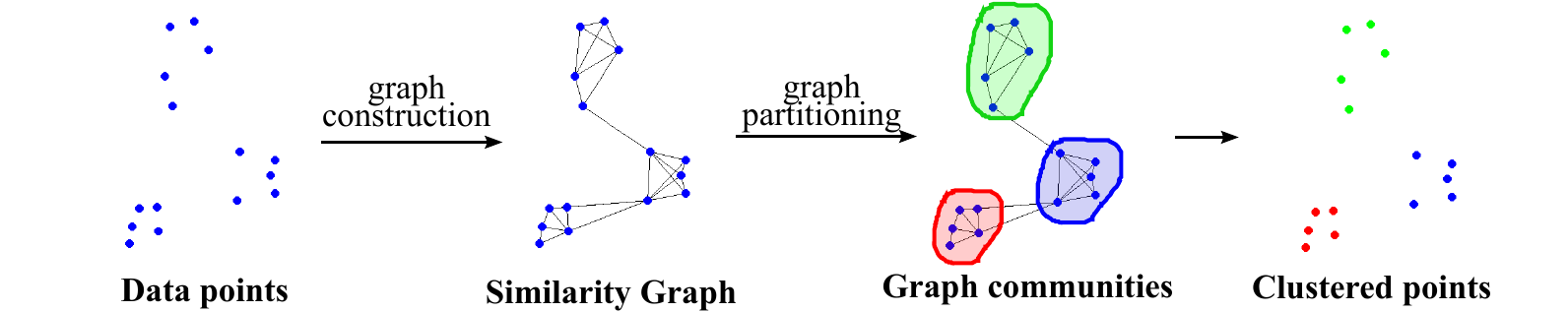}
    \caption{Schematic diagram of the workflow for graph-based clustering}\label{fig:cluster_flow}
\end{figure}

Here, we study the use of multiscale community detection applied to similarity graphs extracted from data for the purpose of unsupervised data clustering.
The basic idea of graph-based clustering is shown schematically in Figure~\ref{fig:cluster_flow}. 
Specifically, we focus on the problem of assessing how to construct graphs that appropriately capture the structure of the dataset with the aim of being used within a multiscale graph-based clustering framework. 
In particular, we carry out an empirical study of different graph construction methods used in conjunction with Markov Stability (MS), a dynamics-based framework for multiscale community detection~\cite{delvenne2010stability,delvenne2013stability}.
As a dynamics-based framework~\cite{delvenne2013stability,lambiotte2014random,delvenne2013stability}, MS provides a unified framework for many multiscale community detection algorithms, such as the RB Potts model~\cite{reichardt2006statistical}, the constant Potts model~\cite{traag2011narrow} and the absolute Potts model~\cite{ronhovde2010local}, and allows for the unsupervised community detection at different levels of resolution. MS has also been shown to allow for the detection of both clique-like and nonclique-like communities in graphs~\cite{schaub2012markov}. MS has been applied successfully to a variety of problems, including protein structures~\cite{delmotte2011protein,amor2014uncovering}, airport networks~\cite{lambiotte2014random}, social networks~\cite{beguerisse2014interest} and neuronal network analyses~\cite{bacik2016flow}. Other dynamical processes have been extensively applied in network analysis, such as temporal networks~\cite{petri2014temporal}, crowded networks~\cite{asllani2018hopping} and network classification~\cite{tran2019scale}. 

In this paper, we evaluate several geometric graph constructions, from methods that use only local distances to others that balance local and global measures, and find that the recently proposed Continuous $k$-nearest Neighbours (CkNN) graph~\cite{berry2016consistent} performs well for graph-based data clustering via community detection. 
We then show how the multiscale capabilities of the Markov Stability to scan across scales 
%framework to show that the identify partitions at different resolutions, and  
%the constructed graph is analysed in different resolutions and the number of clusters can be identified by looking for robust partitions. More importantly, robust clustering results can be obtained when 
can be exploited to deliver robust clusterings, reducing the sensitivity to the parameters of the graph construction. In other words, a range of parameters in the graph construction lead to good clustering performance. 
%which largely reduces the difficulty of finding a proper parameter for graph construction. 
We validate our graph-based clustering approach on real datasets and compare its performance to several other popular clustering methods, including k-means, mixture models, spectral clustering and hierarchical clustering~\cite{rokach2005clustering}.

%\section*{Graph-based clustering}

The rest of the paper is structured as follows. We first introduce several methods for graph construction, apply them to eleven public datasets with ground truths, and evaluate the performance of graph-based data clustering on the ensuing similarity graphs. We then describe briefly the Markov Stability framework for multiscale community detection, and use a synthetic example dataset to illustrate how the multi-resolution clustering reduces the sensitivity to graph construction parameters. Finally, we validate the Markov Stabiity graph-based clustering through comparisons with other clustering methods on real datasets.

%In the next section we compare the different graph construction methods for graph-based clustering with an empirical study on several public datasets and we use a synthetic dataset to illustrate how to achieve multi-resolution clustering with Markov Stability.

\section*{Graph construction methods for data clustering}
\label{section:graph}
%\subsection*{Graph construction methods}
Let us consider a dataset consisting of $n$ samples $\{\mathbf{y}_i\}_{i=1}^n$, where each sample $\mathbf{y}_i$ is a $d$-dimensional vector. 
We will assume that we can define a measure of pairwise dissimilarity between the samples: $d(i,j) \geq 0$. In some cases, the vectors will be defined in a metric space, and the dissimilarity will be a true distance $d(i,j) = ||\mathbf{y}_i-\mathbf{y}_j|| \geq 0$. Other dissimilarity measures, such as the cosine distance, can also be used depending on the application. In the examples below, we will restrict ourselves to Euclidean distances as the measure of dissimilarity.

The high dimensionality of data usually leads to complex and non-linear geometries associated with datasets, posing challenges to standard clustering methods.
The aim of transforming the data into a graph is to capture the complex geometry of the data through the graph topology~\cite{tenenbaum2000global}, so as to reveal the structure of the data via graph-theoretical concepts and tools from complex network analysis.
%or through processes taking place on the graph. 
There exist a variety of ways to construct a graph from a high-dimensional dataset, invoking different principles.
Here we focus on geometric graphs and examine two of the most widely used methods ($\epsilon$-ball graph, $k$-nearest neighbour (kNN) graph) and three recent methods (continuous $k$-nearest neighbours (CkNN) graph~\cite{berry2016consistent}, perturbed minimum spanning tree (PMST)~\cite{zemel2004proximity} and relaxed minimum spanning tree (RMST)~\cite{beguerisse2013finding}). 
These methods can be broadly ascribed to two categories depending on how they use the geometry of the data, as follows.

\subsection*{Neighbourhood based methods: $\epsilon$-ball, kNN and CkNN graphs} 
The idea of neighbourhood based methods is to connect two nodes if they are local neighbours, as given by their pairwise distance $d(i,j)$. 
The two simplest and most popular ways to construct a graph from pairwise distances are the $\epsilon$-ball graph and the $k$-nearest neighbour graph (kNN): in the $\epsilon$-ball graph, any two points at a distance smaller than $\epsilon$ are connected; in the kNN graph, every point is connected to its $k$-th nearest neighbours. These two methods capture the local information of the data but are highly sensitive to the parameters $\epsilon$ or $k$~\cite{maier2008influence}. The parameter is usually set according to the density of the data points but in many datasets, the data points are not uniformly distributed.

A recently proposed method that can resolve this problem is the continuous $k$-nearest neighbours (CkNN) graph~\cite{berry2016consistent}. If $d(i,j)$ is the distance between sample $i$ and sample $j$, and $d^k(i)$ is the distance between sample $i$ and its $k$-th nearest neighbour, the CkNN graph is constructed by connecting sample $i$ and sample $j$ if 
\begin{equation}
    d(i,j) < \delta \sqrt{d^k(i)d^k(j)},
\end{equation}
where $\delta$ is a positive parameter that controls the sparsity of the graph.
Through this construction, the topology of the CkNN graph captures the geometric features of the data with the additional consistency that the CkNN graph Laplacian converges to the Laplace-Beltrami operator in the limit of large %\MB{do you mean large data density or large $n$} 
data~\cite{berry2016consistent}. It should be kept in mind that in CkNN the distance $d(i,j)$ must be a metric to ensure geometrical consistency, whereas kNN graphs can be generated from any dissimilarity measure (i.e., one does not need a true distance) since only the ranking of node closeness matters. 

\subsection*{Minimum spanning tree based methods: PMST and RMST graphs}

A different class of approaches for graph construction attempt to capture the global geometry of the overall dataset by constructing graphs based on measures of global connectivity of the ensuing graph. A popular way to ensure such global connectivity is through the minimum spanning tree (MST)~\cite{cormen2001introduction}, as follows. If we consider the matrix of all pairwise distances $d(i,j)$ as the adjacency matrix of a weighted, fully connected graph, the MST is the subgraph such that all the nodes are path connected and the sum of edge weights is minimised. In other words, the MST provides a graph that connects all the points in the dataset with minimal \textit{global} distance. MST-based approaches can thus capture the geometry of in-homogeneously sampled data points in a high-dimensional space since the MST contains not only local but also global features of the dataset.

In its simplest form, the MST is sometimes added to sparse neighbourhood graphs as a means to guarantee global connectivity of the dataset, i.e., the final graph is the union of the MST and a kNN graph with a small $k$. These schemes, which are sometimes referred as MST+kNN graphs, are the ones we adopt by default in our neighbourhood constructions. 
However, the global properties of the MST can be exploited to generate MST-based graphs from data with distinct properties.
We have explored here the use of two such MST-based algorithms: the perturbed minimum spanning tree (PMST)~\cite{zemel2004proximity} and the relaxed minimum spanning tree (RMST)~\cite{beguerisse2013finding,vangelov14}.

In PMST~\cite{zemel2004proximity}, each data point $\mathbf{y}_i$ is perturbed by a small amount of noise of standard deviation $s_i = rd^k(i)$ ($r\in [0,1]$) where $d^k(i)$ is used as an estimation of the local noise and $r$ is a parameter controlling the level of noise. The MST is then computed for each realisation of the perturbed data and the process is performed repeatedly to generate an ensemble of MSTs. The PSMT graph is given by the union of all the perturbed MSTs, plus the original MST. The intuition behind this algorithm is that random perturbations of the points in high-dimensional space will induce changes inhomogeneously in different parts of the MST, depending on how globally important certain edges of the graph are, i.e., globally important edges will be consistently captured across all MSTs in the perturbed ensemble. One limitation of this algorithm is the heavy computational demand, since both the distance matrix and MST need to be computed for each random realisation. The computational burden makes it impractical to sweep over the parameters of the PMST, so we fix $r=0.5$ and $k=1$ in this paper.

RMST~\cite{vangelov14} proposes a different heuristic for estimating an MST-based graph with lower computational cost. Note that any two nodes are connected by a single path in the MST and we denote the longest edge on this path as $ d^{\mathrm{max}}_{\mathrm{path}(i,j)}$. 
In RMST, the samples $\mathbf{y}_i$ and $\mathbf{y}_j$ are connected if 
\begin{equation}
    d(i,j) < d^{\mathrm{max}}_{\mathrm{path}(i,j)} + \gamma \, \left(d^k(i) + d^k(j)\right),
    \label{eq:RMST}
\end{equation}
where $\gamma>0$ is a parameter that weights the local density (measured by the average distance to the $k$-th neighbours of $\mathbf{y}_i$ and $\mathbf{y}_j$ ) against a global property (the maximum distance found on the MST path linking the samples $i$ and $j$). Similarly to the parameter $\delta$ in CkNN, the value of $\gamma$ controls the sparsity of the resulting graph. The RMST construction was proposed as a means to reconstruct data that have been inhomogeneously sampled from continuous manifolds, and has been shown to provide good description of datasets when preserving a measure of continuity (due to temporal or parametric changes) is important~\cite{beguerisse2013finding}.
%considers both the global feature of the data, which is captured by the longest edge on the MST path in the first term, and the local density, which is taken into account by the second term.

\begin{figure}[!ht]
\centering
  \includegraphics[width=0.7\textwidth]{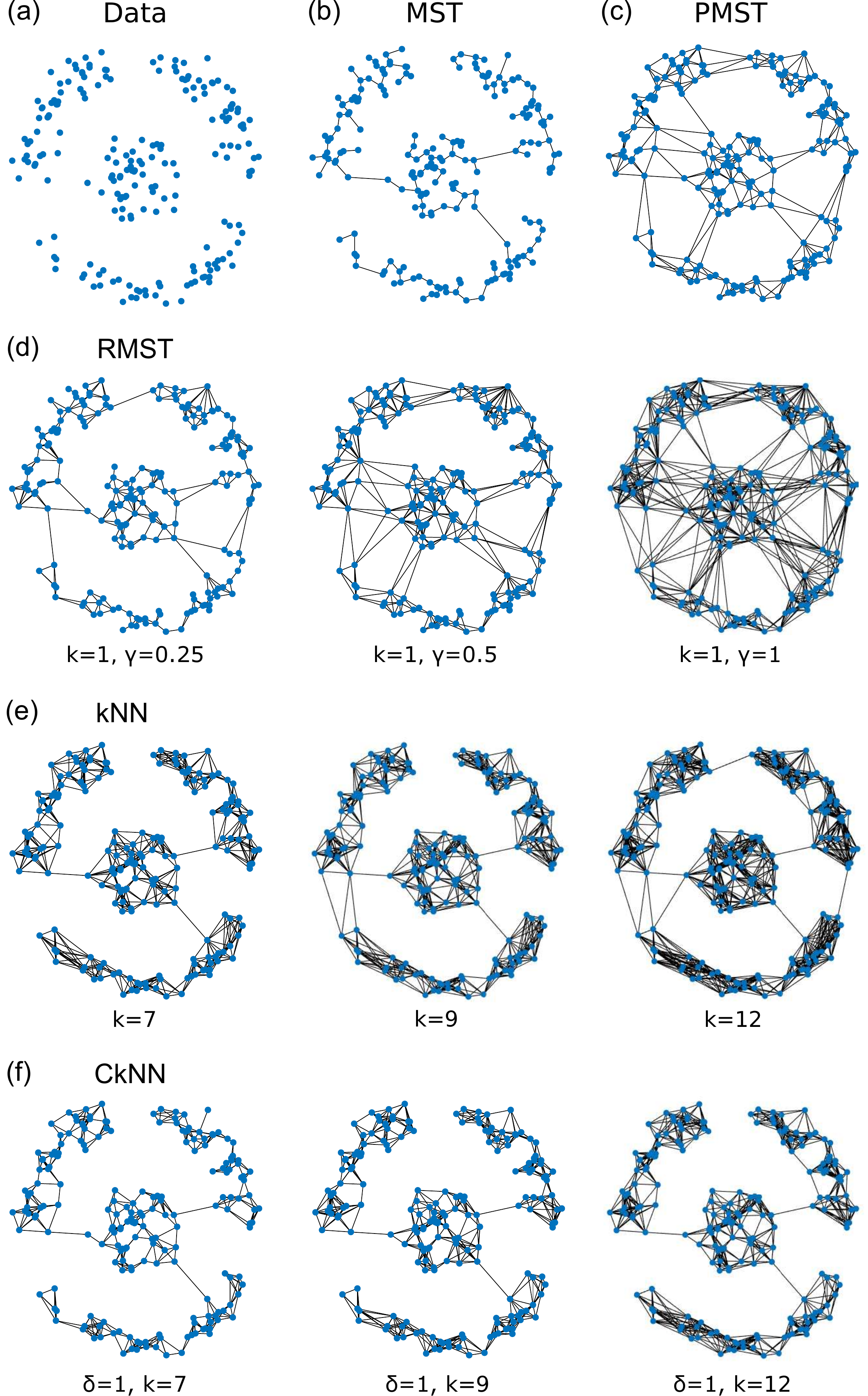}
  \caption{Illustration of the differences introduced by graph construction methods. (a) Synthetic dataset consisting of four groups of randomly sampled points: three groups are sampled from a circle and one group from its centre. (b) The minimum spanning tree (MST) of the dataset based on Euclidean distances. We also  present the graphs constructed with MST-based methods: (c) PMST and (d) RMST,  as well as neighbourhood methods: (e) kNN and (f) CkNN. RMST, kNN and CkNN are shown for different parameters that vary the sparsity of the resulting graphs.}
  \label{fig:five_methods}
\end{figure}

%\MB{HERE --- Maybe move this bit here to finish this section... We do not do anything with this dataset for clustering, so we can just use this for illustration, right?}
In Figure~\ref{fig:five_methods}, we use a synthetic dataset of four groups of points sampled from a geometric structure (a noisy circle and its centre) to illustrate the different graph construction schemes~\cite{ben2001support}. 
The graph representations allow us to gain intuition about the suitability of the different graph constructions for clustering, and the effect of their parameters. For instance, the sparsity of the RMST graph is controlled by the parameter $\gamma$. Note that RMST gives a similar graph to PMST with much lower computational cost. Note also that, for the same number of neighbours $k$, the CkNN gives a sparser graph than the kNN. To make CkNN and kNN comparable, we thus fix the parameter $\delta=1$ in CkNN and vary $k$. 

As discussed above, the MST-based methods are not optimised for clustering but aimed at \textit{manifold learning}.
It follows from~\eqref{eq:RMST} that, in the RMST (and PMST) graphs it is more likely for an edge to appear between node $i$ and node $j$ if the longest edge on the MST path between node $i$ and $j$ is large. This feature makes the geodesic distance on the graph a good approximation to the true distance in the underlying space. Hence both RMST and PMST are closer to \textit{manifold learning} approaches such as Isomap~\cite{tenenbaum2000global}).  
In contrast, the kNN and CkNN graphs tend to have better modular structures and hence appear as potentially more suitable for graph partitioning. We compare these issues in detail in Section~\nameref{section:benchmark} below.

\section*{Markov Stability for graph-based clustering}
\label{section:ms}
Let us consider a graph representing the dataset. The unweighted and undirected graph with $n$ nodes representing the dataset is encoded by the adjacency matrix $A$, where $A_{ij}=1$ if there is an edge connecting node $i$ and $j$ and $A_{ij}=0$ otherwise. The degree of the nodes is summarised in the degree vector $\mathbf{d}$ where $d_i = \sum_j^n A_{ij}$, and we also define the diagonal degree matrix $D$ where $D_{ii} = d_{i}$. The total number of edges of the network is $m = \sum_{i,j} A_{ij}/2$. We then apply multiscale community detection to extract relevant subgraphs in an unsupervised manner using the framework of Markov Stability.

\subsection*{Multiscale community detection with Markov Stability}
Markov Stability is a quality measure for community detection which adopts a dynamical perspective to unfold relevant structures in the graph at all scales as revealed by a diffusion process~\cite{lambiotte2008arXiv,lambiotte2014random, delvenne2010stability,schaub2012markov}. Consider a continuous-time Markov process on the graph governed by the dynamics $\mathbf{\dot{p}} = -\mathbf{p}(I-M)$ where $\mathbf{p}$ is an $n$-dimensional row vector defined on the nodes and $M = D^{-1}A$ is the one step random walk transition matrix. For this Markov process, there is a unique stationary distribution $\vecs{\pi} = \mathbf{d}^T/2m$. Let us denote the autocovariance matrix of this process as $B(t) = \Pi P(t) - \vecs{\pi}^T\vecs{\pi}$ where $\Pi = D/2m $ encodes the stationary distribution and $P(t) = \mathrm{exp}(-t(I-M))$ is the transition matrix. Given a partition $g$ of the nodes into $c$ non-overlapping groups denoted by $g=\{g_1,g_2,...,g_c\}$, the Markov Stability of $g$ is defined as:
\begin{equation}
 r(t,g) = \sum_{s=1}^{c} \sum_{i,j \in g_s} B(t)_{ij}.
 \label{eq:MS}
\end{equation}
A partition has a high value of~\eqref{eq:MS} if the probability of finding a random walker at time $t$ within the group where it started at $t=0$ is higher than that expected by mere chance. In this sense, Markov Stability is a quality function for partitions of a graph and the objective is therefore to find the partitions that achieve high values of the Markov Stability as a function of $t$:
\begin{equation}
 r^*(t) = \max_{g} r(t,g) \quad \text{achieved by the partition $g^*(t)$}.
 \label{eq:maxim_MS}
\end{equation}
The (time) parameter $t$ is the so-called Markov time, and can be understood as the resolution parameter that leads to multiscale community detection~\cite{delvenne2013stability,schaub2012markov}.%---as $t$ grows, the partitions of the graph become coarser.
For small $t$, the number of detected communities is large and the communities capture the local information of the graph. As $t$ becomes larger, there are fewer communities and the communities are able to capture the global features of the graph.
Computationally, the Markov Stability is optimised at different Markov times through a version of the Louvain algorithm~\cite{blondel2008fast}.

%It is worth to mention that the Markov Stability framework can also be used for weighted networks. The modularity is a special case of Markov Stability if we use the one-step random walk transition matrix $D^{-1}A$ instead of the time dependent transition matrix 

%\subsection*{Find the robust partitions}

Through the computational maximisation~\eqref{eq:maxim_MS}, Markov Stability detects optimised partitions $g^*(t)$ at all scales, parameterised by the value of $t$. 
However, we are interested in finding robust partitions and robust scales, in the sense that a partition is found to optimise MS over a long interval of Markov time. We thus compute the dissimilarity between the obtained partitions at different times $t$ and $t'$: 
\begin{align}
    VI(t,t')=VI(g^*(t),g^*(t')),
    \label{eq:VI_tt'}
\end{align}
where we use the variation of information ($VI$)~\cite{meilua2003comparing} as the metric of dissimilarity between partitions. 
If $g^*(t)$ is a robust partition, the partition $g^*(t')$ found at a Markov time $t'$ close to $t$ should be very similar, and hence $VI(t,t')$ will be small. 
%
%The right scale of a partition does not appear at an isolated Markov time but exists in a range of Markov time. At each Markov time, we run the Louvain algorithms $n_L$ times and output the partition that maximises the Markov Stability, which we denote as $g(t)$. 
%If $g(t)$ is a robust partition, the partition $g(t')$ found at another Markov time $t'$ close to $t$ should be identical to $g(t)$. Thus we can compute a pairwise VI matrix across the Markov times. If there are $n_T$ Markov times $t_1,\ldots,t_{n_T}$, we can get a $n_T\times n_T$ matrix $VI(t,t')$ with $VI(t,t')_{ij}= VI(g_{t_i},g_{t_j})$. A zero entry in the matrix $VI(t,t')$ means that the optimal partitions at Markov time $t_i$ and $t_j$ are identical. When scanning across the Markov time $t$, the $VI(t,t')$ matrix will be block-diagonal of zeros.
We therefore look for large diagonal blocks of small values in the $VI(t,t')$ matrix. Such blocks correspond to a robust scale with an associated robust partition. 
%means that for a relatively long time, the same partition appears as the optimal, suggesting that this is a robust partition. Apart from the $VI(t,t')$ matrix, the number of communities as a function of time may also be used as an indicator of robustness. But note that the same number of communities does not mean the same partition and we should use it together with the $VI(t,t')$ matrix. Across time, a long plateau of the number of communities and a big diagonal block of zeros in the $VI(t,t')$ matrix allow us to detect significant scales and corresponding partitions.

As an additional feature, we look for optimised partitions that are also robust to the Louvain optimisation. 
%The first is the robustness of the partitions at a particular Markov time. The Louvain algorithm is used to obtain the graph partition at different Markov times. 
Since the Louvain method is a greedy algorithm dependent on the random initialisation, the consistency of the output of the algorithm can be used as an indicator of the robustness of the solution. At each $t$, we run the Louvain optmisation multiple times and if the Markov time corresponds to a robust scale, the output partition should be always the same.  
%As a metric for partitions, the variation of information (VI)~\cite{meilua2003comparing} is used to measure the consistency. 
%Assume that we repeat the Louvain algorithms $n_L$ times at a Markov time $t$ and get $n_L$ partitions $g_1,\ldots,g_{n_L}$. We can define the
Therefore we expect a low value of the average variation of information of the optimised partitions at time $t$ 
\begin{align}
    VI(t) = \frac{1}{n_L(n_L-1)}\sum_{s=1}^{n_L} \sum_{s'=1}^{n_L} VI(g_s(t),g_{s'}(t)),
    \label{eq:VI_t}
\end{align}
where the Louvain algorithm is run $n_L$ times.
Depending on the structure of the graph, several such robust scales and associated graph partitions might be found, which can then be used as the basis of unsupervised data clustering. 
%right scale and find a good partition in real datasets without the ground truth? 
%The question can be answered by the robustness of the partitions assessed in two aspects.

\subsection*{Using Markov Stability for data clustering}
\begin{figure}[!ht]
  \centering
\includegraphics[width=0.8\textwidth]{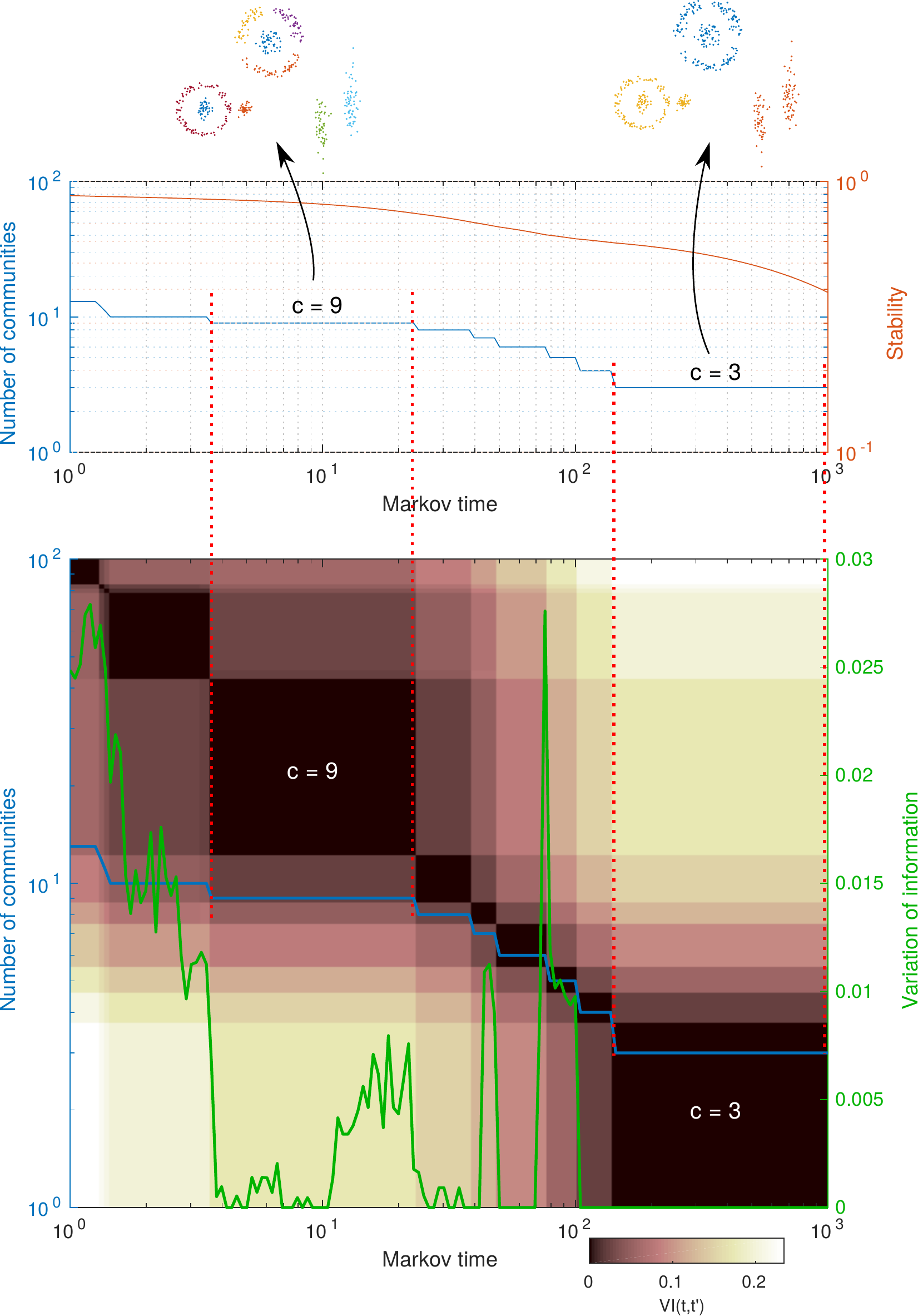}
  \caption{Graph-based unsupervised analysis using Markov Stability of a synthetic dataset with multiscale geometric structure. We generate a CkNN graph ($k=7$, $\delta=1.8$) and scan across Markov time from 1 to 1000. Two significant scales corresponding to robust partitions are identified based on the zero blocks of the $VI(t,t')$ matrix, the long plateaux in the number of communities, and the low values of $VI(t)$. Note that with Markov Stability, there is a range of $\delta$ for the CkNN graph ($k=7$) that can reveal the multiscale structure of the data (see Fig.~\ref{fig:cknn}). }\label{fig:ms}
\end{figure}

We illustrate the application of MS to data clustering through the synthetic dataset in Figure~\ref{fig:ms}. The example dataset has geometric struture and is designed to have two scales, so that it can be divided into 3 big clusters or 9 small clusters. 
First, we construct an unweighted CkNN graph ($k=7$, $\delta=1.8$) and apply MS as described above. We optmise the Markov Stability~\eqref{eq:MS} for $n_T$ Markov times $t\in [1, 1000]$, and at each $t$, we run the Louvain algorithm $n_L=500$ times. For each Markov time, we record 
%the $VI(t)$ for these 500 partitions is computed 
the partition with the maximal Markov Stability, $g^*(t)$, and the average dissimilarity of the partitions found in the $n_L$ optimisations, $VI(t)$~\eqref{eq:VI_t}. 
Once the scan across Markov time is completed, we also compute $VI(t,t')$, the matrix recording the dissimilarity of the optimal partitions found across the scan. 

The results are presented in Figure~\ref{fig:ms}, where, as a function of Markov time $t$, we plot the number of communities in the optimal partition $g^*(t)$; the the optimised Markov Stability $r^*(t)$~\eqref{eq:MS}; the average dissimilarity due to algorithmic variability $VI(t)$; and the dissimilarity of partitions across time given by the $VI(t,t')$ matrix.  The diagonal blocks of low values of $VI(t,t')$ (which also correspond to plateaux in the number of communities) and the low values (or dips) of $VI(t)$ suggest that there are two relevant scales in this graph, which correspond to a finer partition into 9 groups (at small $t$) and  a partition into 3 groups (at larger $t>200$). The inset shows that the partition recovers the planted groups of this synthetic example.

\begin{figure}[!ht]
\centering
  \includegraphics[width=0.8\textwidth]{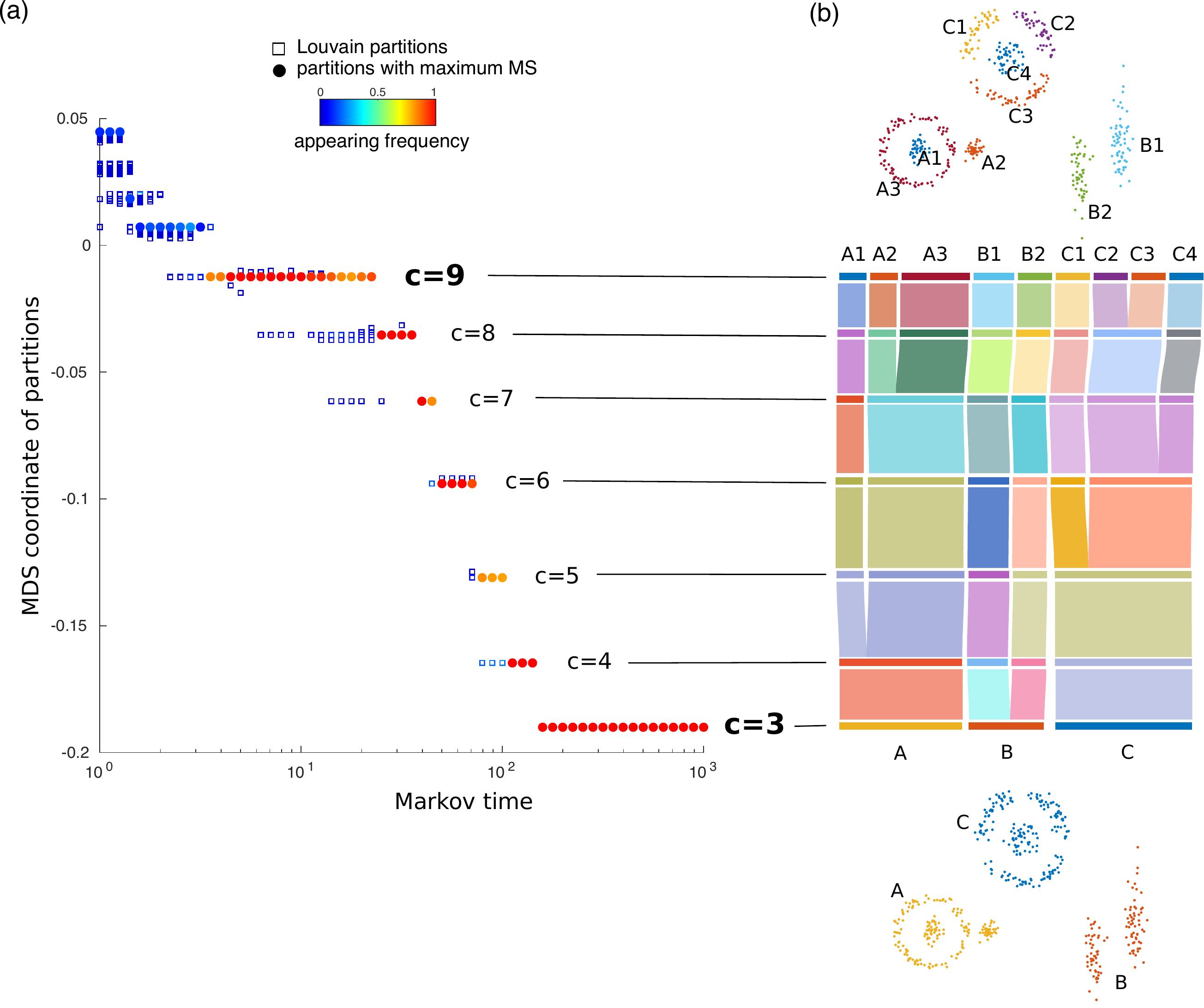}
  \caption{Robustness of the partitions found through Markov Stability for the CkNN graph with $k=7$ and $\delta=1.8$ of the dataset in Figure~\ref{fig:ms}. (a) We show the one-dimensional MDS embedding (based on the $VI$ distance) of all the graph partitions found by the $n_L$ Louvain optimisations for each of the $n_T$ Markov times. The partition with the maximal Markov Stability at each Markov time is labelled by a filled circle, whereas the non-optimal partitions are represented by open squares. The graph partitions are coloured according to the frequency of their appearance in the optimisations.  (b) The Sankey diagram represents the relationship between the partitions found by MS across Markov time, from $c=9$ to $c=3$. Note the hierarchical structure which is not pre-imposed by the algorithm but emerges from the intrinsic properties of the dataset.}\label{fig:mds}
\end{figure}

The robustness of the partitions found across scales is further examined in Figure~\ref{fig:mds}.
%The partition with the maximal Markov Stability at each Markov time is used to compute the $VI(t,t')$ matrix, which shows the relationship between the best partitions across Markov time. 
To understand how the graph partitions evolve with $t$, we compute the $VI$ metric between all the partitions found across the Markov time scan ($n_L \times n_T$) and project them on a low dimensional space using multidimensional scaling (MDS). 
%record the partitions from all $n_L$ Louvain runs at each Markov time and get an ensemble of $n_L \times n_T$ partitions across all Markov times. We use the VI as a distance measure for the partitions and embed the partitions into a low dimensional space with multidimensional scaling (MDS). 
In Figure~\ref{fig:mds}, we use the first MDS coordinate of each of the partitions as a function of $t$ coloured by its number (frequency) of appearances out of the $n_L$ Louvain runs at each Markov time. Several partitions can coexist at a given Markov time, but our numerics show that  the two robust partitions ($c=9$ and $c=3$) have a long-lived high frequency of appearance when the $t$ matches the corresponding resolution. Between $c=9$ and $c=3$, other partitions of lesser robustness appear through mergers of clusters one by one, as shown by the Sankey diagramme in Figure~\ref{fig:mds}, and only exist for short Markov times until the robust partition of $c=3$ appears.

One advantage of using community detection for data clustering is the computational efficiency of fast community detection algorithms~\cite{fortunato2010community}. Empirically, the Louvain algorithm scales nearly linearly with the size of the dataset~\cite{blondel2008fast}. In its full form, Markov Stability has been applied to community detection in networks of sizes up to tens of thousands of nodes~\cite{delmotte2011protein,altuncu2019}. To give an indication, running a full MS scan for a dataset of 2310 samples sweeping 100 Markov times with 100 Louvain runs at each Markov time on a desktop~\footnote{With a 4-core 3.4GHz CPU and MATLAB implementations} requires about 20 minutes. The space complexity of the graph-based clustering is dominated by the storage of the graph, i.e., similar to spectral clustering if the adjacency matrix is used. 
For very large datasets($>20000$), the computational complexity is dominated by the matrix exponential in $P(t)$. The use of linearised (approximate) version of MS allows to scale its use up to graphs with hundreds of thousands of nodes with reduced CPU times, at the cost of some reduction in the quality of the clusterings~\cite{delvenne2013stability}.
The linearised versions also allows the storage of the graph in a sparse matrix to reduce the space complexity.

\subsection*{Scanning across Markov time reduces the sensitivity to graph construction}

Graph-based clustering performance is sensitive to the parameters of the graph construction method, which modulate sparsity, but there is no easy way to select the best parameter if the ground truth is unknown. In practice, the parameters are usually set empirically with little guidance that the chosen parameter will lead to good clustering results for a particular dataset. 
Within the Markov Stability framework, we can use the robustness provided by scanning across Markov time to reduce the sensitivity to the details of graph construction, thus improving the reliability of the detected clusters. 

\begin{figure}[!ht]
    \centering
    \includegraphics[width=0.7\textwidth]{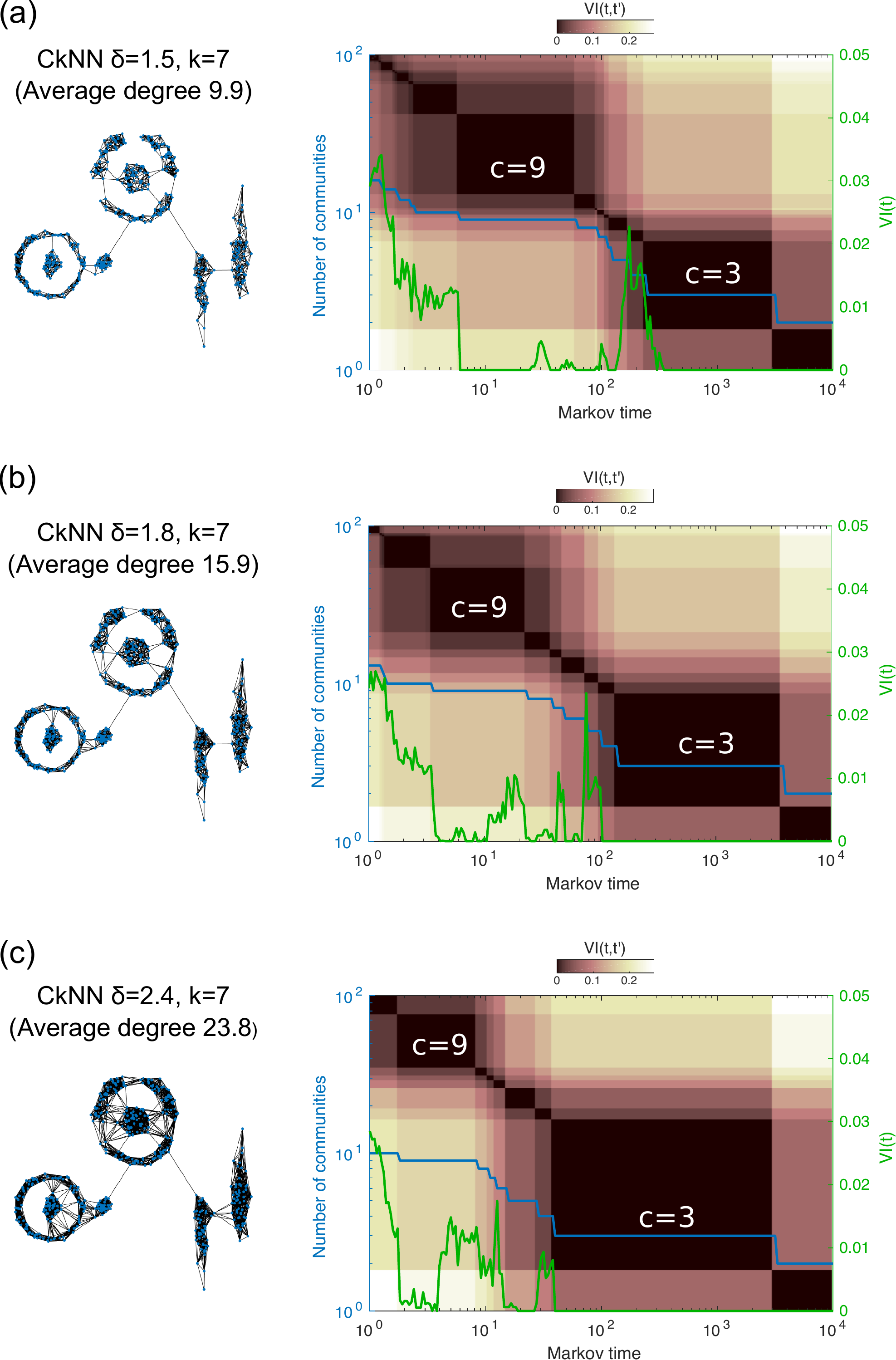}
    \caption{The multiscale character of Markov Stability alleviates the sensitivity to the parameters of graph construction. Using the same synthetic dataset as in Figure~\ref{fig:ms}, MS finds the robust partitions and scales in the data ($c=3$ and $c=9$) in an unsupervised manner for CkNN graphs constructed with a range sparsities as given by varying the parameters: $k=7$ and $\delta=1.5, 1.8, 2.4$ in (a), (b) and (c), respectively. Markov time, as a resolution parameter, allows the community algorithm to reveal the local and global properties of the graph constructed from the data. 
    }\label{fig:cknn}
\end{figure}

To illustrate this idea, We use the same dataset in Figure~\ref{fig:ms} and construct CkNN ($k=7$) graphs with different values of $\delta=1.5, 1.8, 2.4$. Our numerics show that Markov Stability detects the relevant underlying scales of the data ($c=9$ and $c=3$) for the different values of $\delta$, as shown by the long diagonal blocks of low $VI(t,t')$ and the low values of $VI(t)$ in  Figure~\ref{fig:cknn}. 
%presents three cases when $\delta=1.5$, $\delta=1.8$ and $\delta=2.4$. 
Although the degree of the CkNN graph varies markedly with the parameter $\delta$, the two significant scales are identified by scanning across the Markov time. 
Hence the scanning across scales inherent to multiscale community detection provides additional robustness to the parameters of the graph construction algorithm. We also carry out a similar analysis of the clusterings for two real datasets (‘WBDC’ and ‘Control charts’). The clusterings remain robust when varying $k$ in CkNN (see Figure S1 and Figure S2).

\section*{Tests on benchmark real datasets}
\label{section:benchmark}
We have tested several graph-based clustering approaches (both graph constructions and clustering methods) using eleven benchmark datasets from the UCI repository (see Table~\ref{tab:ucidata} for a summary of attributes)~\cite{dua2017}. All the datasets have ground truth labels, which we use to validate the results of the different methods.

\subsection*{Comparison between graph constructions} % on real datasets}

Starting from the Euclidean distance $d(i,j)=||\mathbf{y}_i- \mathbf{y}_j ||_2$,
we generated geometric graphs from each of the eleven UCI datasets using the five graph construction methods described in Section~\nameref{section:graph}. If the constructed graph is disconnected, the MST is added to the graph to ensure a connected graph.
Each graph was analysed using Markov Stability to obtain optimised partitions at any scale, and we selected the closest of those partitions to the ground truth, as measured by the normalised mutual information (NMI)~\cite{strehl2002cluster}. 
The computed NMI is a quality index for the graph construction under clustering.
We also compute the adjusted Rand index (ARI)~\cite{hubert1985comparing} as an additional quality index.  

\begin{table}[!ht]
\centering
\caption{Attributes of the eleven real datasets from the UCI repository used in this study.}

\begin{tabular}{|l|c|c|c|c|}
\hline
 Dataset  & No. of samples, $n$ & Dimension, $d$ & No. of classes, $c^*$ \\ \hline
 Iris  &  150 & 4 & 3  \\
 Glass &  214 & 9 & 6 \\
 Wine  &  178 & 13 & 3 \\
 WBDC  &  569 & 30 & 2  \\
 Control chart & 600 & 60 & 6  \\
 Parkinsons & 195 & 22 & 2 \\
 Vertebral & 310 & 6 & 3 \\
 Breast tissue & 106 & 9 & 6 \\
 Seeds & 210 & 7 & 3 \\
 Image Segmentation & 2310 & 19 & 7 \\
 Yeast & 1484 & 8 & 10 \\\hline
\end{tabular}
\label{tab:ucidata}
\end{table}

The results of the comparison are shown in Table~\ref{tab:nmi_graph}. The kNN and $\epsilon$-ball graphs, both widely-used in many machine learning and both based on local neighbourhoods, give good results for a range of $k$. (Note that $\epsilon$ is set to be the average of the distances to the 7-th neighbour, $d^7(i)$.)
For the MST-based methods, RMST achieves better performance for sparser graphs (with smaller $\gamma$) when the cluster structure is not obscured by the objective of manifold reconstruction (Fig.~\ref{fig:five_methods}). The same applies to the PMST graph. The empirical tests show that the CkNN graph gives the best average results over the eleven datasets for $k=7$ and above. Hence we adopt the CkNN graph with $k$ in the range of 7 to 12 as a good choice for graph-based clustering.

\begin{table}[!ht]
\centering
\caption[Comparison of graph constructions for clustering performance on real datasets]{Comparison of graph constructions in terms of clustering performance (NMI and ARI) on eleven UCI datasets. A high NMI (and ARI) indicate that the best partition found by Markov Stability is similar to the ground truth, i.e., better clustering. The best performance for each dataset is marked in bold.}
%\noindent % needed if the tabularx environment isn't encased in a table environment
\begin{scriptsize}
\begin{tabular}{|@{}p{1.5cm}|@{}p{0.9cm}@{}|@{}p{0.9cm}@{}|@{}p{0.8cm}@{}|@{}p{1.0cm}@{}|@{}p{1.1cm}@{}|@{}p{0.9cm}@{}|@{}p{0.9cm}@{}|@{}p{0.9cm}@{}|@{}p{0.9cm}@{}|@{}p{0.9cm}@{}|@{}p{0.9cm}@{}|}
\multicolumn{12}{c}{\begin{tabular}[c]{@{}c@{}} 
\textbf{Normalised Mutual Information (NMI)}
\end{tabular}}    
\\
\hline
\multirow{2}{*}{Dataset} & \multirow{2}{*}{PMST} & \multirow{2}{*}{$\epsilon$-ball} & \multicolumn{3}{c|}{RMST $k=1$}     & \multicolumn{3}{c|}{kNN} & \multicolumn{3}{c|}{CkNN $\delta=1$} \\ \cline{4-12} 
     &       &         & $\gamma$$=$$0.5$ & $\gamma$$=$$0.25$ & $\gamma$$=$$0.125$ & $\,k$$=$$3$   & $\,k$$=$$7$    & $\,k$$=$$12$  & $\,k$$=$$3$   & $\,k$$=$$7$    & $\,k$$=$$12$  \\ \hline
Iris                     & 0.7764                & 0.8057                        & 0.7106   & 0.7764    & 0.7400       & 0.7627  & 0.8057  & 0.8057 & \textbf{0.8226}  & 0.7980   & 0.7777  \\ \hline
Glass                    & 0.3886                & 0.3862                        & 0.3656   & 0.3646    & 0.3953     & 0.3863  & 0.3927  & 0.3626 & 0.3517  & 0.3941  & \textbf{0.4170}   \\ \hline
Wine                     & 0.8080                 & 0.7972                        & 0.7389   & 0.8400      & \textbf{0.8633}     & 0.7955  & 0.8336  & 0.8215 & 0.7528  & 0.8347  & 0.8113  \\ \hline
WBDC                     & 0.6042                & 0.5819                        & 0.5839   & 0.6042    & 0.6182     & 0.6188  & 0.5944  & 0.6121 & 0.5822  & \textbf{0.7231}  & 0.6113  \\ \hline
Control chart       & 0.8272                & 0.8404                        & 0.7602   & 0.7133    & 0.8130      & 0.8266  & 0.8520   & 0.852  & 0.8078  & \textbf{0.8883}  & 0.8531  \\ \hline
Parkinson                & 0.2220                 & 0.2065                        & 0.2012   & 0.2143    & 0.2811     & \textbf{0.3175}  & 0.2815  & 0.2176 & 0.2113  & 0.2973  & 0.2423  \\ \hline
Vertebral                & 0.5043                & 0.6273                        & 0.5432   & 0.5323    & 0.5999     & 0.6060   & 0.5928  & 0.5424 & \textbf{0.6450}   & 0.6018  & 0.6116  \\ \hline
Breast tissue            & 0.5722                & 0.5616                        & 0.5298   & \textbf{0.5820}     & 0.5461     & 0.5447  & 0.5525  & 0.5243 & 0.5253  & 0.5648  & 0.5601  \\ \hline
Seeds                    & 0.7293                & 0.6943                        & 0.7361   & \textbf{0.7570}     & 0.6810      & 0.7515  & 0.7458  & 0.7318 & 0.6384  & 0.7381  & 0.7245  \\ \hline
Image Seg.       & 0.6948              & 0.6605                        & 0.6762   & 0.7072     & \textbf{0.7488}      & 0.6154  & 0.6465  & 0.6667 & 0.6012  & 0.6347  & 0.6541  \\ \hline
Yeast            & 0.2881                & 0.3051                        & 0.2952   & 0.2626     & 0.2563      & 0.2764  & 0.2997  & \textbf{0.3080} & 0.2473  & 0.2959  & 0.3072  \\ \hline\hline
\textbf{Average}      & 0.5835                & 0.5879                        & 0.5583   & 0.5776    & 0.5945     & 0.5910  & 0.5997  & 0.5859 & 0.5623   & \textbf{0.6155}  & 0.5973  \\ \hline
\end{tabular}
\\
\bigskip

\begin{tabular}{|@{}p{1.5cm}|@{}p{0.9cm}@{}|@{}p{0.9cm}@{}|@{}p{0.8cm}@{}|@{}p{1.0cm}@{}|@{}p{1.1cm}@{}|@{}p{0.9cm}@{}|@{}p{0.9cm}@{}|@{}p{0.9cm}@{}|@{}p{0.9cm}@{}|@{}p{0.9cm}@{}|@{}p{0.9cm}@{}|}
%\multicolumn{12}{c}{ARI}   
\multicolumn{12}{c}{\begin{tabular}[c]{@{}c@{}} 
\textbf{Adjusted Rand Index (ARI)}
\end{tabular}}  
\\
\hline
\multirow{2}{*}{Dataset} & \multirow{2}{*}{PMST} & \multirow{2}{*}{$\epsilon$-ball} & \multicolumn{3}{c|}{RMST $k=1$}     & \multicolumn{3}{c|}{kNN} & \multicolumn{3}{c|}{CkNN $\delta=1$} \\ \cline{4-12} 
     &       &         & $\gamma$$=$$0.5$ & $\gamma$$=$$0.25$ & $\gamma$$=$$0.125$ & $\,k$$=$$3$   & $\,k$$=$$7$    & $\,k$$=$$12$  & $\,k$$=$$3$   & $\,k$$=$$7$    & $\,k$$=$$12$  \\ \hline
Iris                     & 0.7420                 & 0.7592                        & 0.6603   & 0.7420     & 0.6957     & 0.7191  & 0.7592  & 0.7592 & \textbf{0.8184}  & 0.7455  & 0.7445  \\ \hline
Glass                    & 0.2323                & 0.2099                        & 0.1983   & 0.2029    & 0.2258     & 0.2278  & 0.2231  & 0.2266 & 0.2134  & 0.2398  & \textbf{0.2496}  \\ \hline
Wine                     & 0.8350                 & 0.8072                        & 0.7375   & 0.8712    & \textbf{0.8823}     & 0.8025  & 0.8498  & 0.8349 & 0.7414  & 0.8471  & 0.8360   \\ \hline
WBDC                     & 0.7193                & 0.7114                        & 0.7014   & 0.7193    & 0.7369     & 0.7368  & 0.7010   & 0.7310  & 0.6697  & \textbf{0.8244}  & 0.7200    \\ \hline
Control chart & 0.6929                & 0.7364                        & 0.5694   & 0.5371    & 0.6991     & 0.6748  & 0.6824  & 0.7071 & 0.6902  & \textbf{0.8280}   & 0.7227  \\ \hline
Parkinson        & 0.2267                & 0.2205                        & 0.2038   & 0.1540     & 0.2556     & 0.2176  & 0.2001  & 0.2045 & 0.1165  & \textbf{0.2667}  & 0.2101  \\ \hline
Vertebral                & 0.5257                & \textbf{0.6445}                        & 0.5702   & 0.5411    & 0.6015     & 0.5982  & 0.5802  & 0.5330  & 0.6441  & 0.6113  & 0.6302  \\ \hline
Breast tissue            & \textbf{0.4689}                & 0.4494                        & 0.4100     & 0.4017    & 0.3494     & 0.4012  & 0.4272  & 0.4078 & 0.3631  & 0.3764  & 0.4471  \\ \hline
Seeds                    & 0.7353                & 0.7497                        & 0.7458   & 0.7589    & 0.6687     & 0.7876  & \textbf{0.7889}  & 0.7761 & 0.6402  & 0.7764  & 0.7655  \\ \hline
Image Seg.                    & 0.6060&	0.5144&	0.6030&	\textbf{0.6193}&	0.5942&	0.3800	&0.4669	&0.5471&	0.3791	&0.4522&	0.5121  \\ \hline
Yeast                    & 0.1908&	0.2368&	0.1827&	0.1772&	0.1649&	0.1755&	0.2230&	\textbf{0.2531}&	0.1797&	0.1942&	0.2294\\ \hline\hline
\textbf{Average}     & 0.5432&	0.5490&	0.5075&	0.5204&	0.5340	&0.5201&	0.5365&	0.5437	&0.4960&	\textbf{0.5602}&	0.5516  \\ \hline
\end{tabular}
\end{scriptsize}

\label{tab:nmi_graph}
\end{table}

%The reason why the CkNN graph has a good performance for clustering analysis is partly due to the fact that the CkNN graph can provide a better discrete approximation of the diffusion operators than other graph constructions on the underlying data manifold. 

The CkNN graph is constructed by using a variable bandwidth diffusion kernel~\cite{berry2016variable} where the bandwidth is inversely proportional to the sampling density and allows for uniform estimation errors over the underlying manifold, while the graph construction which uses a fixed bandwidth kernel will have large errors in areas of small sampling density. By such a construction, the graph Laplacian of the CkNN graph converges to the the Laplacian operator in the manifold. This explains why the CkNN graph shows a better clustering performance than the other graph construction when the graph is partitioned with Markov Stability, which considers a diffusion process on the graph.

%The reason why CkNN graph has a good performance for clustering analysis is partly due to the fact that the CkNN graph can provide a better discrete approximation of the diffusion operators than other graph constructions on the underlying data manifold, and many community detection methods are closely related to diffusion or random walks, including the Markov Stability framework, which will be introduced in the next section. The CkNN graph is constructed by using a variable bandwidth diffusion kernel~\cite{berry2016variable} where the bandwidth is inversely proportional to the sampling density and allows for uniform estimation errors over the underlying manifold, while the graph construction which uses a fixed bandwidth kernel will have large errors in areas of small sampling density. By such a construction, the graph Laplacian of the CkNN graph converges to the the Laplacian operator in the manifold. 

\subsection*{Comparison between clustering methods} % on real datasets}

In this section, we evaluate the performance of graph-based clustering through Markov Stability against several other clustering methods applied to the datasets from the UCI repository.
We include a variety of clustering approaches. Model-based methods include k-means and Gaussian mixture (clustering is repeated 50 times and partition with the best objective function is reported). We also apply hierarchical clustering 
%which is widely used due to its simplicity and visualisation power, the 
with complete linkage and Euclidean distance is used as the distance measure. Since graph-based clustering is closely related to spectral clustering, we compare to two spectral clustering methods: the multiclass n-cut algorithm~\cite{yu2003multiclass} and the classic NJW algorithm~\cite{ng2002spectral}. The affinity matrix for these two spectral clustering algorithms is calculated with a local density kernel as described in~\cite{zelnik2005self}. 
Note that all these algorithms need the number of clusters to be given as an input. Hence we use the number of classes in the ground truth $c^*$ as an input to set the number of clusters. For comparability, in the case of Markov Stability, we construct the CkNN graph with Euclidean distance ($k=7$ and $\delta=1$) and find the optimised partition with the number of clusters equal to $c^*$. %\MB{I am confused about this... Are you fixing $c=c^*$ also for MS in this first numerical experiment? I thought so but then you had something different in the caption of the figure...}

The results are presented in Table~\ref{tab:nmi_compare}. Given the distinct features of the datasets, no method is expected to achieve consistently better performance across all datasets. The hierarchical clustering performs worst, as it is easily affected by the noise in real datasets. Model-based methods, such as the Gaussian mixture model, can achieve good performance if the properties of the data fit the assumptions (e.g., the Iris dataset), but can also perform poorly. On datasets that have non-convex geometries, graph-based methods and spectral clustering tend to perform better. On average, the Markov Stability approach achieves the best NMI and ARI scores.

\begin{table}[!ht]
\centering
\caption{Comparison of performance of clustering methods on eleven real UCI datasets.
The number of clusters is fixed to the number of clusters of the ground truth and given as an input.}

\begin{tabular}{|l|c|c|c|c|c||c|}
\multicolumn{7}{c}{\begin{tabular}[c]{@{}c@{}} \textbf{Normalised Mutual Information (NMI)}\\ (Number of clusters fixed to ground truth: $c =c^*$)\end{tabular}
}    \\ 
\hline
Dataset & 
k-means & 
\begin{tabular}[c]{@{}c@{}}Gaussian  \\ mixture\end{tabular} &
Hierarchical &
\begin{tabular}[c]{@{}c@{}}Spectral  \\ N-cut\end{tabular} & \begin{tabular}[c]{@{}c@{}}Spectral  \\ NJW\end{tabular} & \begin{tabular}[c]{@{}c@{}}MS on \\ CkNN graph  \end{tabular} \\ 
\hline
%\multicolumn{7}{c}{NMI}      \\ \hline
Iris  ($c^*= 3$)  & 0.7582  & \textbf{0.8997}   & 0.7221   & 0.7578   & 0.7665  & 0.7980 \\ \hline
Glass ($c^*= 6$) & 0.3163  & \textbf{0.4033}   & 0.1548  & 0.2899   & 0.3175    & 0.3932 \\ \hline
Wine ($c^*= 3$)    & 0.8759  & 0.9120   & 0.6144    & \textbf{0.9276}   & \textbf{0.9276}   & 0.8347 \\ \hline
WBDC ($c^*= 2$)   & 0.5546    & 0.5275  & 0.0102 & 0.5434  & 0.5565            & \textbf{0.7231} \\ \hline
Control chart ($c^*= 6$) & 0.6907   & 0.5348  & 0.7070  & 0.7924  & 0.7891   & \textbf{0.8489} \\ \hline
Parkinson ($c^*= 2$)  & 0.0970   & 0.0387  & 0.0519  & \textbf{0.2189}  & \textbf{0.2189}  & 0.1334 \\ \hline
Vertebral  ($c^*= 3$)  & 0.4120  & 0.5559 & 0.1280   & 0.3969  & 0.3901   & \textbf{0.5971} \\ \hline
Breast tissue ($c^*= 6$)   & 0.5446   & \textbf{0.5487}  & 0.4745     & 0.5481  & 0.5366  & 0.5291  \\ \hline
Seeds  ($c^*= 3$)  & 0.7279  & \textbf{0.8111}   & 0.7010  & 0.6881  & 0.6881           & 0.7142 \\ \hline
Image Seg.  ($c^*= 7$)  & 0.5840&	0.5697&	0.0359&	0.5929&	\textbf{0.6466}          & 0.5816 \\ \hline
Yeast  ($c^*= 10$)  & \textbf{0.2955}&	0.2071&	0.1846&	0.2794&	0.2801   & 0.2909 \\ \hline\hline
\textbf{Average} &  0.5324 &	0.5462 &	0.3440	&0.5487&	0.5561   & \textbf{0.5858} \\ \hline
\end{tabular}
\\
\bigskip
\begin{tabular}{|l|c|c|c|c|c||c|}
\multicolumn{7}{c}{\begin{tabular}[c]{@{}c@{}} \textbf{Adjusted Rand Index (ARI)}\\ (Number of clusters fixed to ground truth: $c = c^*$)\end{tabular}}    \\
\hline
Dataset & 
k-means & 
\begin{tabular}[c]{@{}c@{}}Gaussian  \\ mixture\end{tabular} &
Hierarchical &
\begin{tabular}[c]{@{}c@{}}Spectral  \\ N-cut\end{tabular} & \begin{tabular}[c]{@{}c@{}}Spectral  \\ NJW\end{tabular} & \begin{tabular}[c]{@{}c@{}}MS on \\ CkNN graph \end{tabular} \\ 
\hline
Iris  ($c^*= 3$) & 0.7302    & \textbf{0.9039}   & 0.6423  & 0.7430   & 0.7570            & 0.7455 \\ \hline
Glass ($c^*= 6$) & 0.1722   & \textbf{0.2649}   & 0.0350   & 0.1104  & 0.1483       & 0.2380 \\ \hline
Wine ($c^*= 3$)  & 0.8975     & 0.9325   & 0.5771    & \textbf{0.9471}   & \textbf{0.9471}   & 0.8471 \\ \hline
WBDC ($c^*= 2$)  & 0.6707   & 0.6436  & 0.0048    & 0.6661  & 0.6777   & \textbf{0.8244} \\ \hline
Control chart ($c^*= 6$) & 0.5207      & 0.2699    & 0.5374   & 0.6835  & 0.6803      & \textbf{0.7056} \\ \hline
Parkinson ($c^*= 2$)  & -0.0978    & -0.0502   & -0.0710                  & 0.1608   & 0.1608    & \textbf{0.2659} \\ \hline
Vertebral  ($c^*= 3$) & 0.3147   & 0.5256    & -0.0327    & 0.2986   & 0.2888      & \textbf{0.6096} \\ \hline
Breast tissue ($c^*= 6$)    & 0.2876   & 0.2872   & 0.1895    & \textbf{0.3983}   & 0.3806     & 0.3659 \\ \hline
Seeds  ($c^*= 3$) & 0.7733    & \textbf{0.8167}    & 0.6863  & 0.7189  & 0.7189  & 0.7432 \\ \hline 
Image Seg.  ($c^*= 7$) & 0.4605	&0.4522&	0.0011&	0.3360&	\textbf{0.4849}  & 0.4150 \\ \hline
Yeast  ($c^*= 10$) & 0.1658 &	0.0972&	0.1083&	0.1483	&0.1481 & \textbf{0.1746} \\ \hline\hline
 \textbf{Average}     & 0.4450 &	0.4676	&0.2435&	0.4737	&0.4902 &    \textbf{0.5395} \\ \hline
\end{tabular}
\\
\label{tab:nmi_compare}
\end{table}

To further validate the quality of the robust partitions found by Markov Stability, we also carried out MS clustering in a fully unsupervised manner, i.e., without providing the number of classes in the ground truth as an input. Using the principles described in Section~\nameref{section:ms}, we identify robust scales and robust partitions in order to establish the number of clusters inherent to the data in an unsupervised manner. The number of clusters detected by MS and the clustering performances in terms of NMI and ARI are presented in Table~\ref{tab:num_cluster_compare}, together with the results obtained in Table~\ref{tab:nmi_compare}, where the number of clusters in the ground truth was provided. Although the number of detected clusters differs slightly from the ground truth, the clusters found in the unsupervised MS have comparable ARI values and a higher average NMI value which indicates that the clusters are of good quality and provide more information about the ground truth. This highlights the capability of Markov Stability as an unsupervised data clustering approach in practice where the number of clusters are usually unknown.

\begin{table}[!ht]
\centering
\caption{Graph-based clustering via Markov Stability performs equally well in a totally unsupervised manner, even if the number of clusters is not given \textit{a priori}. }

\begin{tabular}{|l|c|c|}
%\multicolumn{3}{c}{NMI}      \\ \hline
\multicolumn{3}{c}{\begin{tabular}[c]{@{}c@{}} \textbf{Normalised Mutual Information (NMI)}\end{tabular}}    \\

\hline
Dataset & \begin{tabular}[c]{@{}c@{}}Markov Stability\\ ($c$ fixed to ground truth $c^*$) \end{tabular} & \begin{tabular}[c]{@{}c@{}}Markov Stability\\  ($c$ found by MS unsupervised)\end{tabular} \\ \hline
Iris   & 0.7980 ($c^*= 3$)   & \textbf{0.7980} ($c=3$) \\ \hline
Glass   & \textbf{0.3932} ($c^*= 6$)  & 0.3687 ($c=5$)\\ \hline
Wine    & 0.8347 ($c^*= 3$)  & \textbf{0.8347} ($c=3$)\\ \hline
WBDC    & {0.7231} ($c^*= 2$)           & \textbf{0.7231} ($c=2$) \\ \hline
Control chart& 0.8489 ($c^*= 6$)  & \textbf{0.8520} ($c=4$) \\ \hline
Parkinson    & 0.1334 ($c^*= 2$) & \textbf{0.2973} ($c=3$) \\ \hline
Vertebral     & 0.5971 ($c^*= 3$) & \textbf{0.6018} ($c=2$) \\ \hline
Breast tissue    & \textbf{0.5291} ($c^*= 6$) & 0.5104 ($c=3$)  \\ \hline
Seeds     & 0.7142 ($c^*= 3$) & \textbf{0.7142} ($c=3$)\\ \hline
Image Seg.     & 0.5816 ($c^*= 7$) & \textbf{0.6338} ($c=13$)\\ \hline
Yeast     & 0.2909 ($c^*= 10$) & \textbf{0.2909} ($c=10$)\\ \hline\hline 
\textbf{Average}  & 0.5858   & \textbf{0.6023} \\ \hline
%\multicolumn{3}{c}{}   \\
%\multicolumn{3}{c}{ARI}      \\
\end{tabular}

\bigskip 

\begin{tabular}{|l|c|c|}
\multicolumn{3}{c}{\begin{tabular}[c]{@{}c@{}} \textbf{Adjusted Rand Index (ARI)}\end{tabular}}    \\
\hline
Dataset & \begin{tabular}[c]{@{}c@{}}Markov Stability\\ ($c$ fixed to ground truth $c^*$) \end{tabular} & \begin{tabular}[c]{@{}c@{}}Markov Stability\\  ($c$ found by MS unsupervised)\end{tabular} \\ \hline
Iris         & 0.7455 ($c^*= 3$) & \textbf{0.7455} ($c=3$)\\ \hline
Glass        & \textbf{0.2380} ($c^*= 6$) & 0.2086 ($c=5$)\\ \hline
Wine         & 0.8471 ($c^*= 3$) & \textbf{0.8471} ($c=3$)\\ \hline
WBDC         & 0.8244 ($c^*= 2$) & \textbf{0.8244} ($c=2$)\\ \hline
Control chart& \textbf{0.7056} ($c^*= 6$)& 0.6824 ($c=4$)\\ \hline
Parkinson    & 0.2659 ($c^*= 2$)& \textbf{0.2667} ($c=3$)\\ \hline
Vertebral    & 0.6096 ($c^*= 3$)& \textbf{0.6113} ($c=2$)\\ \hline
Breast tissue& 0.3659  ($c^*= 6$)& \textbf{0.3764} ($c=3$) \\ \hline
Seeds        & 0.7432 ($c^*= 3$)& \textbf{0.7432} ($c=3$)\\ \hline
Image Seg.        & 0.4150 ($c^*= 7$)& \textbf{0.4516} ($c=13$)\\ \hline
Yeast        & 0.1746 ($c^*= 10$)& \textbf{0.1746} ($c=10$)\\ \hline
\hline
\textbf{Average}  & \textbf{0.5395} & {0.5393} \\ \hline
\end{tabular}
\label{tab:num_cluster_compare}
\end{table}

% \begin{table}[!ht]
% \caption{The number of clusters detected by the Markov Stability analysis on the real datasets.}
% \centering
% \begin{tabular}{|l|c|c|}
% \hline
%  Dataset  & 
%  \begin{tabular}[c]{@{}l@{}} No. of clusters detected\\by Markov Stability\end{tabular}
%   & 
%  \begin{tabular}[c]{@{}l@{}}No. of clusters in\\ground truth \end{tabular}  \\ \hline
%  Iris  &  3  & 3 \\ \hline
%  Glass &  5  & 6 \\\hline
%  Wine  &  3  & 3 \\\hline
%  WBDC  &  2  & 2 \\\hline
%  Control chart & 4  & 6 \\\hline
%  Parkinsons & 3  & 2 \\\hline
%  Vertebral & 2 & 3 \\\hline
%  Breast tissue & 3  & 6 \\\hline
%  Seeds & 3  & 3 \\ \hline
% \end{tabular}
% \label{tab:num_cluster}
% \end{table}

\section*{Conclusion}

We have investigated the use of multiscale community detection for graph-based data clustering. The first step in graph-based clustering is to construct a graph from the data, and our empirical study shows that the recently proposed CkNN graph is a good choice for this purpose. In contrast to other neighbourhood-based graph constructions like kNN or $\epsilon$-ball graphs, the CkNN graph is designed to provide a consistent discrete approximation of the diffusion operator on the underlying data manifold. Since many community detection methods are closely related to diffusion or random walks (e.g., Markov Stability and spectral methods), this explains the good performance of CkNN for clustering purposes. Other graph construction methods specifically designed for manifold learning (e.g, RMST) performed well but are not optimised for cluster separation.

Our work has also examined the suitability of multiscale community detection as a means for unsupervised data clustering. Specifically, we have used the Markov Stability framework, which employs a diffusion process on the graph to detect the presence of relevant subgraphs at all scales.   
%is a community detection framework that considers a Markov process on the graph and is able to detect the multiscale structure. After estimating the graph structure from the data, Markov Stability, a framework for multiscale graph partition, is employed to find the community structures of the graph.
%In the Markov Stability framework, 
The time of the diffusion process acts as a resolution parameter and a cost function for graph partitioning is optimised at different scales by scanning time. Robust partitions and robust scales can be identified by analysing the consistency of the ensemble of optimised partitions found by the Louvain algorithm. Our numerics show that the Markov Stability framework is able to determine the number of clusters and reveal the multiscale structure in data. Further, by scanning Markov time, the MS analysis can reduce the sensitivity to the parameters in the graph construction step, thus improving the robustness of graph-based clustering.

We have validated our graph-based clustering approach on several real datasets by comparing with other popular clustering methods, including k-means, Gaussian mixture model, hierarchical clustering, and two spectral clustering algorithms. The graph-based clustering method achieves the best NMI and ARI values on average across the datasets. Importantly, we show that the clustering can be done in a completely unsupervised way (without assuming a knowledge of the number of clusters), whereas for the other standard methods the number of clusters needs to be given as an input.

Our study also suggests several directions of future work. Here we showed that the CkNN graph is a good choice for graph-based clustering, but it will be interesting to establish the performance of CkNN in other data mining problems, such as manifold learning where graphs also play a important role~\cite{yan2007graph}. We showed that the variation of information of partitions and the ensemble of partitions found by greedy optimisation can be used to guide the identification of robust partitions. However, a quantitative, statistically sound process to choose the significant partitions automatically would be desirable and useful in practice. Another interesting direction is the potential study of the outputs of MS clustering using methods from topological data analysis (TDA). Scanning across Markov time results in an ensemble of time-dependent weighted graphs with adjacency matrices $DP(t)$. It would the be possible to use methods from TDA to characterise the persistent homology of the graphs as a function of the Markov time $t$ to detect robust structures in the data across scales and its relationship with the observed hierarchy of clusters of increasing coarseness.
%As shown in Figure~\ref{fig:mds}, the ensembles of partitions found by greedy optimisation may be worth further investigation as it reveals information about the landscape of the Markov Stability function.

%The original modularity~\cite{newman2004finding} can also be seen as a random walk based method, since it is equivalent to the discrete-time Markov Stability at time one where the one-step random walk transition matrix $D^{-1}A$ is used instead of the time dependent transition matrix $P(t)$. Or equivalently, scanning all Markov time is the same as analysing a series of time dependent graphs with the adjacency matrix $DP(t)$. Thus the properties of the graphs $DP(t)$ still need more investigation.

%By sparsifying the similarity/distance matrix of the data, one can obtain an adjacency matrix of an unweighted sparse graph, instead of a fully connected weighted one. A sparse representation might help to form the piece-wise constant eigenvectors~\cite{azran2006spectral}, which are important to find the clusters.

%%%%%%%%%%%%%%%%%%%%%%%%%%%%%%%%%%%%%%%%%%%%%%
%%                                          %%
%% Backmatter begins here                   %%
%%                                          %%
%%%%%%%%%%%%%%%%%%%%%%%%%%%%%%%%%%%%%%%%%%%%%%

\section*{Competing interests}
  The authors declare that they have no competing interests.

\section*{Author's contributions}
  ZL and MB conceived of the idea of the study. ZL implemented the methods and ran the numerical experiments. Both authors wrote, reviewed and approved the manuscript.

\section*{Acknowledgements}
  This work was supported by the European Commission [European Union 7th Framework Programme for research, technological development and demonstration under grant agreement no. 607466], and the Engineering and Physical Sciences Research Council (EPSRC) through grant EP/N014529/1 to M.B..

\newpage
\section*{Supplementary}

\begin{figure}[!ht]
\centering
   \includegraphics[width=0.7\textwidth]{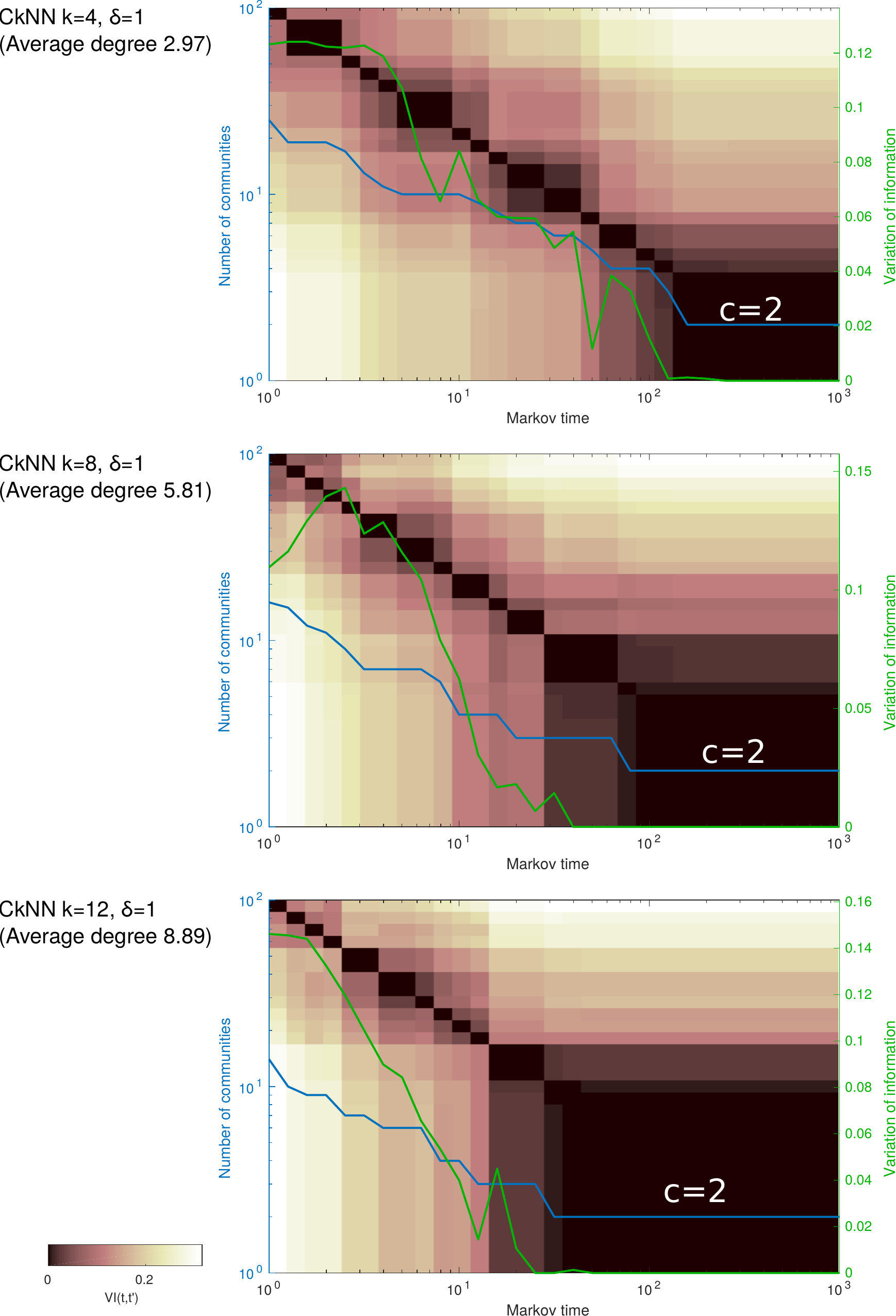}
   \caption{Figure S2: The multiscale Markov Stability analysis of the WBDC dataset as the parameter $k$ in CkNN is varied. The clustering into 2 groups is robust to the change of this parameter of the graph construction.}
\end{figure}

\begin{figure}[!ht]
\centering
   \includegraphics[width=0.7\textwidth]{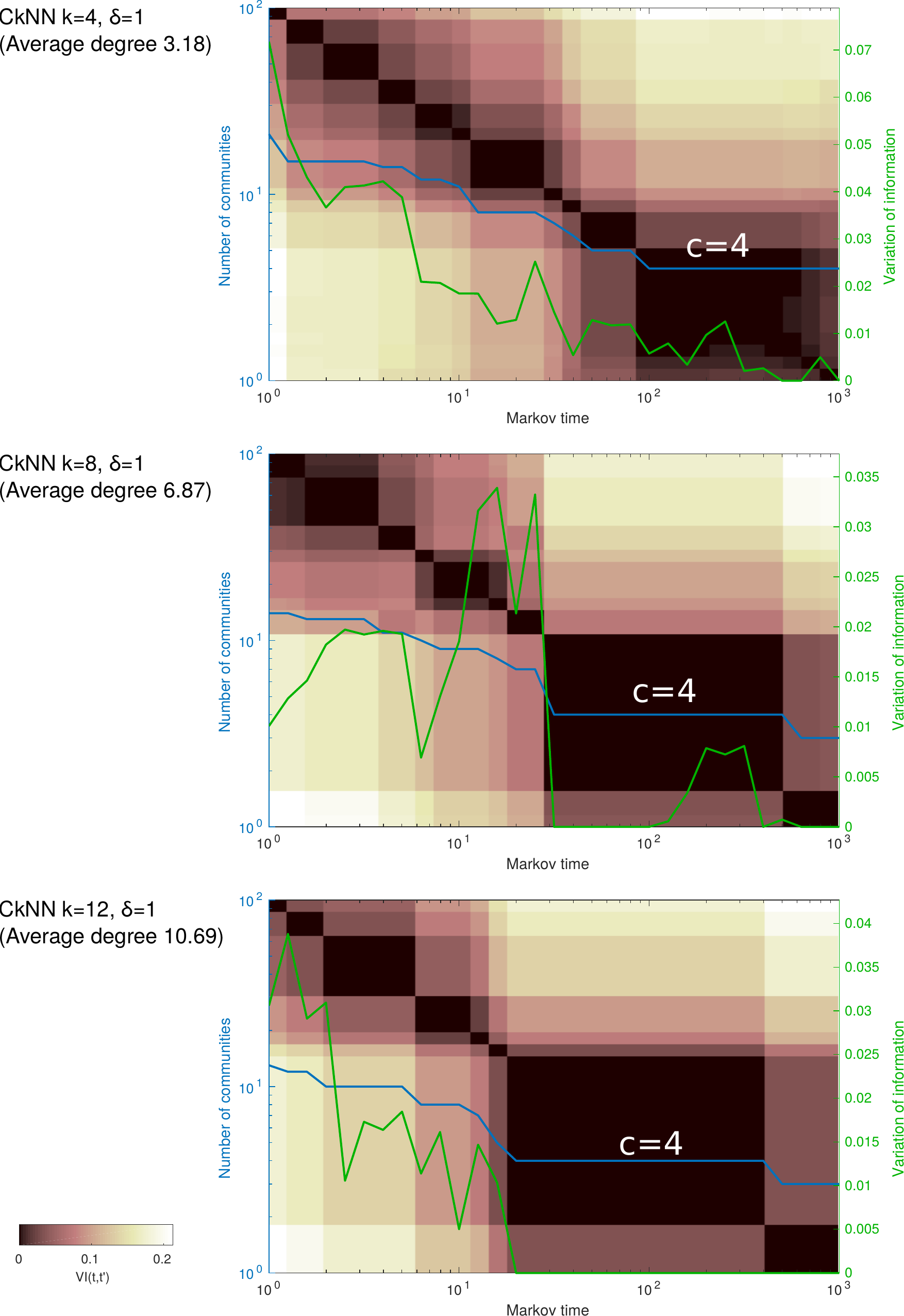}
   \caption{Figure S1: 
   The multiscale Markov Stability analysis of the Control Chart dataset as the parameter $k$ in CkNN is varied. The clustering into 4 groups is robust to the change of this parameter of the graph construction.}
\end{figure}    

\begin{table}[!ht]
\centering
\caption{Table S1: The performance of clustering methods on eleven real UCI datasets measured by Purity. Purity is defined as the proportion of correctly clustered samples (with respect to the groups in the `ground truth').}
\begin{tabular}{|l|c|c|c|c|c|c|c|}
\multicolumn{8}{c}{\begin{tabular}[c]{@{}c@{}} \textbf{Purity} \end{tabular}
}    \\ 
\hline
\multirow{2}{*}{Dataset}
& \multicolumn{6}{c|}{$c$ fixed to ground truth} & $c$ found by MS
\\ \cline{2-8}
&k-means & 
\begin{tabular}[c]{@{}c@{}}Gaussian  \\ mixture\end{tabular} &
Hierarchical &
\begin{tabular}[c]{@{}c@{}}Spectral  \\ N-cut\end{tabular} & \begin{tabular}[c]{@{}c@{}}Spectral  \\ NJW\end{tabular} & \begin{tabular}[c]{@{}c@{}} MS \\ CkNN  \end{tabular} & \begin{tabular}[c]{@{}c@{}} MS \\ CkNN  \end{tabular} \\ 
\hline
Iris   & 0.8933	&\textbf{0.9800}&	0.8400&	0.9000&	0.9067&	0.9000	&0.9000 \\ \hline
Glass   & 0.5467 & 0.5280 & 0.4112 & 0.5327 & 0.5748 & \textbf{0.6636} & 0.6589  \\ \hline
Wine  & 0.9663 & 0.9775 & 0.8371 & \textbf{0.9831} & \textbf{0.9831} & 0.9494 & 0.9494  \\ \hline
WBDC & 0.9104 & 0.9016 & 0.6309 & 0.9086 & 0.9121 & \textbf{0.9543} & \textbf{0.9543}   \\ \hline
Control chart & 0.7000 & 0.5333 & 0.6117 & \textbf{0.8133} & 0.8117 & 0.7983 & 0.6667   \\ \hline
Parkinson     & 0.7538 & 0.7538 & 0.7538 & 0.7538 & 0.7538 & 0.7897 & \textbf{0.8564}   \\ \hline
Vertebral     & 0.6774 & \textbf{0.7774} & 0.4839 & 0.7065 & 0.6968 & 0.7613 & 0.7613   \\ \hline
Breast tissue & 0.5472 & 0.6038 & 0.4717 & 0.6132 & \textbf{0.6321} & 0.6132 & 0.5283   \\ \hline
Seeds   & 0.9190 & \textbf{0.9333} & 0.8762 & 0.9000 & 0.9000 & 0.9048 & 0.9048   \\ \hline
Image Seg.    & 0.5541 & 0.5545 & 0.1671 & 0.5926 & 0.6541 & 0.6082 & \textbf{0.7649}   \\ \hline
Yeast   & \textbf{0.5492} & 0.4623 & 0.3908 & 0.5303 & 0.5283 & 0.5364 & 0.5364   \\ \hline 
\hline
Average & 0.7289 & 0.7278 & 0.5886 & 0.7486 & 0.7594 & 0.7708 & \textbf{0.7710} \\ \hline
\end{tabular}
\end{table}


\begin{thebibliography}{10}

\bibitem{xu2005survey}
R.~Xu and D.~Wunsch, ``Survey of clustering algorithms,'' {\em IEEE
  Transactions on Neural Networks}, vol.~16, no.~3, pp.~645--678, 2005.

\bibitem{sugar2003finding}
C.~A. Sugar and G.~M. James, ``Finding the number of clusters in a dataset: An
  information-theoretic approach,'' {\em Journal of the American Statistical
  Association}, vol.~98, no.~463, pp.~750--763, 2003.

\bibitem{azran2006spectral}
A.~Azran and Z.~Ghahramani, ``Spectral methods for automatic multiscale data
  clustering,'' in {\em Computer Vision and Pattern Recognition, 2006 IEEE
  Computer Society Conference on}, vol.~1, pp.~190--197, IEEE, 2006.

\bibitem{jain1999data}
A.~K. Jain, M.~N. Murty, and P.~J. Flynn, ``Data clustering: a review,'' {\em
  ACM computing surveys (CSUR)}, vol.~31, no.~3, pp.~264--323, 1999.

\bibitem{macqueen1967some}
J.~MacQueen, ``Some methods for classification and analysis of multivariate
  observations,'' in {\em Proceedings of the Fifth Berkeley Symposium on
  Mathematical Statistics and Probability, Volume 1: Statistics}, (Berkeley,
  Calif.), pp.~281--297, University of California Press, 1967.

\bibitem{dempster1977maximum}
A.~P. Dempster, N.~M. Laird, and D.~B. Rubin, ``Maximum likelihood from
  incomplete data via the em algorithm,'' {\em Journal of the royal statistical
  society. Series B (methodological)}, pp.~1--38, 1977.

\bibitem{shi2000normalized}
J.~Shi and J.~Malik, ``Normalized cuts and image segmentation,'' {\em Pattern
  Analysis and Machine Intelligence, IEEE Transactions on}, vol.~22, no.~8,
  pp.~888--905, 2000.

\bibitem{ng2002spectral}
A.~Y. Ng, M.~I. Jordan, and Y.~Weiss, ``On spectral clustering: Analysis and an
  algorithm,'' in {\em Proceedings of the 14th International Conference on
  Neural Information Processing Systems: Natural and Synthetic}, NIPS'01,
  (Cambridge, MA, USA), pp.~849--856, MIT Press, 2001.

\bibitem{de2005spectral}
V.~R. De~Sa, ``Spectral clustering with two views,'' in {\em Proceedings of
  ICML workshop on learning with multiple views}, vol.~2005, pp.~20--27, 2005.

\bibitem{ye2016fuse}
W.~Ye, S.~Goebl, C.~Plant, and C.~B{\"o}hm, ``Fuse: {F}ull spectral
  clustering,'' in {\em Proceedings of the 22nd ACM SIGKDD International
  Conference on Knowledge Discovery and Data Mining}, pp.~1985--1994, ACM,
  2016.

\bibitem{von2007tutorial}
U.~Von~Luxburg, ``A tutorial on spectral clustering,'' {\em Statistics and
  computing}, vol.~17, no.~4, pp.~395--416, 2007.

\bibitem{schaub2019}
M.~T. Schaub, J.-C. Delvenne, R.~Lambiotte, and M.~Barahona, ``Multiscale
  dynamical embeddings of complex networks,'' {\em Phys. Rev. E}, vol.~99,
  p.~062308, Jun 2019.

\bibitem{alpert1999spectral}
C.~J. Alpert, A.~B. Kahng, and S.-Z. Yao, ``Spectral partitioning with multiple
  eigenvectors,'' {\em Discrete Applied Mathematics}, vol.~90, no.~1,
  pp.~3--26, 1999.

\bibitem{dhillon2001co}
I.~S. Dhillon, ``Co-clustering documents and words using bipartite spectral
  graph partitioning,'' in {\em Proceedings of the seventh ACM SIGKDD
  international conference on Knowledge discovery and data mining},
  pp.~269--274, ACM, 2001.

\bibitem{hagen1992new}
L.~Hagen and A.~B. Kahng, ``New spectral methods for ratio cut partitioning and
  clustering,'' {\em IEEE transactions on computer-aided design of integrated
  circuits and systems}, vol.~11, no.~9, pp.~1074--1085, 1992.

\bibitem{chungspectral}
F.~Chung, {\em Spectral Graph Theory}.
\newblock No.~no. 92 in CBMS Regional Conference Series, Conference Board of
  the Mathematical Sciences.

\bibitem{lambiotte2008arXiv}
R.~{Lambiotte}, J.~C. {Delvenne}, and M.~{Barahona}, ``{Laplacian Dynamics and
  Multiscale Modular Structure in Networks},'' {\em arXiv e-prints},
  p.~arXiv:0812.1770, Dec 2008.

\bibitem{lambiotte2014random}
R.~Lambiotte, J.-C. Delvenne, and M.~Barahona, ``Random walks, {Markov}
  processes and the multiscale modular organization of complex networks,'' {\em
  IEEE Transactions on Network Science and Engineering}, vol.~1, no.~2,
  pp.~76--90, 2014.

\bibitem{delvenne2010stability}
J.-C. Delvenne, S.~N. Yaliraki, and M.~Barahona, ``Stability of graph
  communities across time scales,'' {\em Proceedings of the National Academy of
  Sciences}, vol.~107, no.~29, pp.~12755--12760, 2010.

\bibitem{altuncu2019}
M.~T. Altuncu, E.~Mayer, S.~N. Yaliraki, and M.~Barahona, ``From free text to
  clusters of content in health records: an unsupervised graph partitioning
  approach,'' {\em Applied Network Science}, vol.~4, no.~1, p.~2, 2019.

\bibitem{cheng2009learning}
B.~{Cheng}, J.~{Yang}, S.~{Yan}, Y.~{Fu}, and T.~S. {Huang}, ``Learning with
  $\ell ^{1}$-graph for image analysis,'' {\em IEEE Transactions on Image
  Processing}, vol.~19, pp.~858--866, April 2010.

\bibitem{tenenbaum2000global}
J.~B. Tenenbaum, V.~De~Silva, and J.~C. Langford, ``A global geometric
  framework for nonlinear dimensionality reduction,'' {\em Science}, vol.~290,
  no.~5500, pp.~2319--2323, 2000.

\bibitem{beguerisse2013finding}
M.~Beguerisse-D{\'\i}az, B.~Vangelov, and M.~Barahona, ``Finding role
  communities in directed networks using {Role-Based Similarity, Markov
  Stability and the Relaxed Minimum Spanning Tree},'' in {\em Global Conference
  on Signal and Information Processing (GlobalSIP), 2013 IEEE}, pp.~937--940,
  IEEE, 2013.

\bibitem{dhillon2004kernel}
I.~S. Dhillon, Y.~Guan, and B.~Kulis, ``Kernel k-means: spectral clustering and
  normalized cuts,'' in {\em Proceedings of the tenth ACM SIGKDD international
  conference on Knowledge discovery and data mining}, pp.~551--556, ACM, 2004.

\bibitem{kulis2009semi}
B.~Kulis, S.~Basu, I.~Dhillon, and R.~Mooney, ``Semi-supervised graph
  clustering: a kernel approach,'' {\em Machine learning}, vol.~74, no.~1,
  pp.~1--22, 2009.

\bibitem{bronstein2017}
M.~M. {Bronstein}, J.~{Bruna}, Y.~{LeCun}, A.~{Szlam}, and P.~{Vandergheynst},
  ``Geometric deep learning: Going beyond euclidean data,'' {\em IEEE Signal
  Processing Magazine}, vol.~34, pp.~18--42, July 2017.

\bibitem{maier2008influence}
M.~Maier, U.~v. Luxburg, and M.~Hein, ``Influence of graph construction on
  graph-based clustering measures,'' in {\em Proceedings of the 21st
  International Conference on Neural Information Processing Systems}, NIPS'08,
  (USA), pp.~1025--1032, Curran Associates Inc., 2008.

\bibitem{daitch2009fitting}
S.~I. Daitch, J.~A. Kelner, and D.~A. Spielman, ``Fitting a graph to vector
  data,'' in {\em Proceedings of the 26th Annual International Conference on
  Machine Learning}, pp.~201--208, ACM, 2009.

\bibitem{maier2013result}
M.~Maier, U.~Von~Luxburg, and M.~Hein, ``How the result of graph clustering
  methods depends on the construction of the graph,'' {\em ESAIM: Probability
  and Statistics}, vol.~17, pp.~370--418, 2013.

\bibitem{jebara2009graph}
T.~Jebara, J.~Wang, and S.-F. Chang, ``Graph construction and b-matching for
  semi-supervised learning,'' in {\em Proceedings of the 26th Annual
  International Conference on Machine Learning}, pp.~441--448, ACM, 2009.

\bibitem{delvenne2013stability}
J.-C. Delvenne, M.~T. Schaub, S.~N. Yaliraki, and M.~Barahona, ``The stability
  of a graph partition: A dynamics-based framework for community detection,''
  in {\em Dynamics On and Of Complex Networks, Volume 2: Applications to
  Time-Varying Dynamical Systems} (A.~Mukherjee, M.~Choudhury, F.~Peruani,
  N.~Ganguly, and B.~Mitra, eds.), pp.~221--242, New York, NY: Springer New
  York, 2013.

\bibitem{reichardt2006statistical}
J.~Reichardt and S.~Bornholdt, ``Statistical mechanics of community
  detection,'' {\em Physical Review E}, vol.~74, no.~1, p.~016110, 2006.

\bibitem{traag2011narrow}
V.~A. Traag, P.~Van~Dooren, and Y.~Nesterov, ``Narrow scope for
  resolution-limit-free community detection,'' {\em Physical Review E},
  vol.~84, no.~1, p.~016114, 2011.

\bibitem{ronhovde2010local}
P.~Ronhovde and Z.~Nussinov, ``Local resolution-limit-free potts model for
  community detection,'' {\em Physical Review E}, vol.~81, no.~4, p.~046114,
  2010.

\bibitem{schaub2012markov}
M.~T. Schaub, J.-C. Delvenne, S.~N. Yaliraki, and M.~Barahona, ``Markov
  dynamics as a zooming lens for multiscale community detection: non
  clique-like communities and the field-of-view limit,'' {\em PloS one},
  vol.~7, no.~2, p.~e32210, 2012.

\bibitem{delmotte2011protein}
A.~Delmotte, E.~W. Tate, S.~N. Yaliraki, and M.~Barahona, ``Protein multi-scale
  organization through graph partitioning and robustness analysis: application
  to the myosin--myosin light chain interaction,'' {\em Physical biology},
  vol.~8, no.~5, p.~055010, 2011.

\bibitem{amor2014uncovering}
B.~Amor, S.~Yaliraki, R.~Woscholski, and M.~Barahona, ``Uncovering allosteric
  pathways in caspase-1 using markov transient analysis and multiscale
  community detection,'' {\em Molecular BioSystems}, vol.~10, no.~8,
  pp.~2247--2258, 2014.

\bibitem{beguerisse2014interest}
M.~Beguerisse-D{\'\i}az, G.~Garduno-Hern{\'a}ndez, B.~Vangelov, S.~N. Yaliraki,
  and M.~Barahona, ``Interest communities and flow roles in directed networks:
  the {Twitter} network of the {UK} riots,'' {\em Journal of The Royal Society
  Interface}, vol.~11, no.~101, p.~20140940, 2014.

\bibitem{bacik2016flow}
K.~A. Bacik, M.~T. Schaub, M.~Beguerisse-D{\'\i}az, Y.~N. Billeh, and
  M.~Barahona, ``Flow-based network analysis of the {Caenorhabditis} elegans
  connectome,'' {\em PLoS computational biology}, vol.~12, no.~8, p.~e1005055,
  2016.

\bibitem{petri2014temporal}
G.~Petri and P.~Expert, ``Temporal stability of network partitions,'' {\em
  Physical Review E}, vol.~90, no.~2, p.~022813, 2014.

\bibitem{asllani2018hopping}
M.~Asllani, T.~Carletti, F.~Di~Patti, D.~Fanelli, and F.~Piazza, ``Hopping in
  the crowd to unveil network topology,'' {\em Physical review letters},
  vol.~120, no.~15, p.~158301, 2018.

\bibitem{tran2019scale}
Q.~H. Tran, Y.~Hasegawa, {\em et~al.}, ``Scale-variant topological information
  for characterizing the structure of complex networks,'' {\em Physical Review
  E}, vol.~100, no.~3, p.~032308, 2019.

\bibitem{berry2016consistent}
T.~Berry and T.~Sauer, ``Consistent manifold representation for topological
  data analysis,'' {\em Foundations of Data Science}, vol.~1, no.~1, pp.~1--38,
  2019.

\bibitem{rokach2005clustering}
L.~Rokach and O.~Maimon, ``Clustering methods,'' in {\em Data mining and
  knowledge discovery handbook}, pp.~321--352, Boston, MA: Springer US, 2005.

\bibitem{zemel2004proximity}
M.~A. Carreira-Perpi\~{n}\'{a}n and R.~S. Zemel, ``Proximity graphs for
  clustering and manifold learning,'' in {\em Proceedings of the 17th
  International Conference on Neural Information Processing Systems}, NIPS'04,
  (Cambridge, MA, USA), pp.~225--232, MIT Press, 2004.

\bibitem{cormen2001introduction}
T.~H. Cormen, C.~E. Leiserson, R.~L. Rivest, and C.~Stein, {\em Introduction to
  Algorithms, Third Edition}.
\newblock Cambridge, MA: The MIT Press, 3rd~ed., 2009.

\bibitem{vangelov14}
B.~Vangelov, {\em Unravelling Biological Processes using Graph Theoretical
  Algorithms and Probabilistic Models}.
\newblock PhD thesis, Imperial College London, 2014.

\bibitem{ben2001support}
A.~Ben-Hur, D.~Horn, H.~T. Siegelmann, and V.~Vapnik, ``Support vector
  clustering,'' {\em Journal of machine learning research}, vol.~2, no.~Dec,
  pp.~125--137, 2001.

\bibitem{blondel2008fast}
V.~D. Blondel, J.-L. Guillaume, R.~Lambiotte, and E.~Lefebvre, ``Fast unfolding
  of communities in large networks,'' {\em Journal of statistical mechanics:
  theory and experiment}, vol.~2008, no.~10, p.~P10008, 2008.

\bibitem{meilua2003comparing}
M.~Meil{\u{a}}, ``Comparing clusterings by the variation of information,'' in
  {\em Learning Theory and Kernel Machines} (B.~Sch{\"o}lkopf and M.~K.
  Warmuth, eds.), (Berlin, Heidelberg), pp.~173--187, Springer Berlin
  Heidelberg, 2003.

\bibitem{fortunato2010community}
S.~Fortunato, ``Community detection in graphs,'' {\em Physics reports},
  vol.~486, no.~3, pp.~75--174, 2010.

\bibitem{dua2017}
D.~Dheeru and E.~Karra~Taniskidou, ``{UCI} machine learning repository,'' 2017.

\bibitem{strehl2002cluster}
A.~Strehl and J.~Ghosh, ``Cluster ensembles---a knowledge reuse framework for
  combining multiple partitions,'' {\em Journal of machine learning research},
  vol.~3, no.~Dec, pp.~583--617, 2002.

\bibitem{hubert1985comparing}
L.~Hubert and P.~Arabie, ``Comparing partitions,'' {\em Journal of
  classification}, vol.~2, no.~1, pp.~193--218, 1985.

\bibitem{berry2016variable}
T.~Berry and J.~Harlim, ``Variable bandwidth diffusion kernels,'' {\em Applied
  and Computational Harmonic Analysis}, vol.~40, no.~1, pp.~68--96, 2016.

\bibitem{yu2003multiclass}
S.~X. Yu and J.~Shi, ``Multiclass spectral clustering,'' in {\em Proceedings
  Ninth IEEE International Conference on Computer Vision}, pp.~313--319, Oct
  2003.

\bibitem{zelnik2005self}
L.~Zelnik-Manor and P.~Perona, ``Self-tuning spectral clustering,'' in {\em
  Proceedings of the 17th International Conference on Neural Information
  Processing Systems}, NIPS'04, (Cambridge, MA, USA), pp.~1601--1608, MIT
  Press, 2004.

\bibitem{yan2007graph}
S.~Yan, D.~Xu, B.~Zhang, H.-J. Zhang, Q.~Yang, and S.~Lin, ``Graph embedding
  and extensions: A general framework for dimensionality reduction,'' {\em IEEE
  transactions on pattern analysis and machine intelligence}, vol.~29, no.~1,
  pp.~40--51, 2007.

\end{thebibliography}
\end{document}